\documentclass[a4paper, 11pt]{report}
\usepackage[english]{babel}
\usepackage{blindtext}
\usepackage[left=2.5cm,right=3cm, top = 2.5cm, bottom = 3cm]{geometry}
\usepackage{keylistings,keyverbfile, keybook, keyjavadl}
\usepackage{amsthm, amsmath, amssymb}
\usepackage{enumitem}
\usepackage{mdframed}
\usepackage{caption}
\usepackage{multirow}
\usepackage{subcaption}
\usepackage{placeins}
\usepackage{graphicx}

\DeclareSymbolFont{amscal}{OMS}{cmsy}{m}{n}
\SetSymbolFont{amscal}{bold}{OMS}{cmsy}{b}{n}
\DeclareSymbolFontAlphabet{\mathcal}{amscal}

\newtheorem{definition}{Definition}[chapter]
\newtheorem{note}{Note}[chapter]

\newcommand{\KeY}{Ke\kern-0.1emY}
\newcommand{\keyword}[1] {{\textbf{\texttt{#1}}}}

\newcommand{\old}{\keyword{\textbackslash old}}
\newcommand{\result}{\keyword{\textbackslash result}}
\newcommand{\fresh}{\keyword{\textbackslash fresh}}
\newcommand{\jforall}{\textbf{\texttt{\textbackslash forall}}}
\newcommand{\jexists}{\keyword{\textbackslash exists}}
\newcommand{\everything}{\keyword{\textbackslash everything}}
\newcommand{\nothing}{\keyword{\textbackslash nothing}}
\newcommand{\also}{\keyword{also}}
\newcommand{\public}{\keyword{public}}
\newcommand{\specpublic}{\keyword{spec\_public}}
\newcommand{\norm}{\keyword{normal\_behavior}}
\newcommand{\excep}{\keyword{exceptional\_behavior}}
\newcommand{\signals}{\keyword{signals}}
\newcommand{\signalso}{\keyword{signals\_only}}

\newcommand{\nullable}{\keyword{nullable}}
\newcommand{\pure}{\keyword{pure}}
\newcommand{\accessible}{\keyword{accessible}}
\newcommand{\diverge}{\keyword{diverge}}
\newcommand{\decreases}{\keyword{decreases}}
\newcommand{\loopinv}{\keyword{loop\_invariant}}

\newcommand{\ensures}{\keyword{ensures}}
\newcommand{\requires}{\keyword{requires}}
\newcommand{\assignable}{\keyword{assignable}}
\newcommand{\invariant}{\keyword{invariant}}

\newcommand{\eabs}{\keyword{ensures\_abs}}
\newcommand{\rabs}{\keyword{requires\_abs}}
\newcommand{\aabs}{\keyword{assignable\_abs}}

\newcommand{\defabs}{\keyword{def}}

\newcommand{\Types}{\ensuremath{\mathrm{T}}}
\newcommand{\subtype}{\ensuremath{\mathrm{\sqsubseteq}}}
\newcommand{\Fun}{\ensuremath{\mathrm{F}}}
\newcommand{\Pred}{\ensuremath{\mathrm{P}}}
\newcommand{\Vars}{\ensuremath{\mathrm{V}}}
\newcommand{\PVars}{\ensuremath{\mathrm{PV}}}
\newcommand{\sigfunc}{\ensuremath{\mathrm{\alpha}}}
\newcommand{\Prg}{\ensuremath{\mathrm{Prg}}}
\newcommand{\Domain}{\ensuremath{\mathcal{D}}}
\newcommand{\State}{\ensuremath{\mathcal{S}}}
\newcommand{\Interp}{\ensuremath{\mathcal{I}}}
\newcommand{\Kripke}{\ensuremath{\mathcal{K}}}

\begin{document}

\begin{titlepage}

\begin{center}
	\vspace*{3cm}
	{\Huge \rm \bf Realization and Extension\\[0.5ex]
	 of Abstract Operation Contracts\\[0.9ex]
	 for Program Logic}\\[10ex]
	{\Large Maria Pelevina}\\
	{\Large m.pelevina@gmail.com}\\[10ex]
	Bachelor's Thesis\\[10ex]
	\begin{tabular}{ll}	
	Supervisors: 	& Prof. Dr. Reiner H\"{a}hnle\\
			& Dr. Richard Bubel\\
	\end{tabular}
	\vfill
	Computer Science Department\\
	Technische Universit\"{a}t Darmstadt\\
	May, 2014\\
\end{center}

\end{titlepage}

\chapter*{Acknowledgements}
First and foremost I would like to thank Prof. Dr. Reiner H\"ahnle for offering me an opportunity to work on such a thrilling topic that is pushing the edge of science in the field of formal verification.

I am deeply indebted to my supervisor Dr. Richard Bubel for many hours he spent guiding me through all stages of this work. His willingness to help and immediate reaction to any arising questions and problems have ensured the success of my thesis.

Last but not least, I am grateful to my loved ones, whose never-ending support helped me to get through highs and lows of the entire process and, in particular, to my boyfriend, without whose patience and encouragement I would not have finished this thesis.

\chapter*{Abstract}
For engineering software with formal correctness proofs it is crucial that proofs can be efficiently reused in case the software or its specification is changed.
Unfortunately, in reality even slight changes in the code or its specification often result in disproportionate waste of verification effort: For instance, whenever a method's specification is modified and as a consequence the proof of its correctness breaks, all other proofs based on this specification break too.
Abstract method calls is a recently proposed verification rule for method calls that allows for efficient systematic reuse of proofs.
In this thesis, we implement, extend and evaluate this approach within the KeY verification system.

\tableofcontents

\chapter{Introduction}
\section{Motivation}
One of the distinctive features of modern software is that it has changed from static to constantly evolving products. That means frequent changes to accommodate new client needs, to optimize existing functionalities and, of course, to fix discovered bugs. It has also become standard in software industry to assure software's correct behavior with help of testing, which, as it is well known, can only show the presence of bugs and not their absence \cite{Dijkstra69}. While just testing the software may be sufficient for many every-day application scenarios, there are other areas like avionics, automotive or medical engineering, where the correctness of software is a much more critical requirement.

Software verification with formal methods has emerged as a mean to mathematically prove the absence of bugs in a program, or more precisely, that a program functions as intended. However, in the current state of art the costs of formal verification of a non-trivial piece of software are still very high. Although several successful steps have been made on the path of automation of software verification, it is still far from being a push-button technology and often requires interaction by an experienced verification engineer. 

The fact that modern software has become constantly evolving further worsens this situation. Even slight changes in code may result in disproportionate waste of specification and verification effort. For instance, whenever a method implementation is modified and as a consequence the proof of its correctness breaks, then all other proofs based on this one break too. Whereas a naive and straightforward approach in this case would be to start the verification of the affected methods from scratch, we aim at saving the effort by reusing parts of the old proofs that remain unaffected by the changes.

In this thesis we extend the recently proposed approach for effective systematic reuse of proofs \cite{Haehnle13} in the context of the \KeY\ verification system. 

\section{Context}
The \emph{\KeY\ system} is a formal verification environment that uses dynamic logic to deductively prove the correctness of programs written in the Java programming language. It's specification methodology follows the \emph{design by contract} approach suggested by Meyer \cite{Meyer92} in which the notion of program correctness is expressed through contracts -- mutual obligations binding program components. Relevant components of an object-oriented program being class methods, a behavior of a method \texttt{m} is captured in form of a \emph{contract} with a \emph{precondition} $P$ and a \emph{postcondition} $E$. The intuitive meaning is similar to that of a Hoare triple \{$P$\}\texttt{m}\{$E$\}: if the method m is called in a state satisfying the precondition $P$, then, assuming termination and determinism of the programing language, the postcondition $E$ should hold in the resulting state after the execution of \texttt{m}.

\KeY\ translates the source code and the provided \emph{method contracts} into formulas of \emph{dynamic logic} \cite{Harel00}, also called \emph{proof obligations}. More precisely, the \KeY\ system uses \emph{Java Dynamic Logic (JavaDL)} which is an extension of the first-order predicate logic with modal operators for handling executable code fragments written in the \emph{Java Card} language.

The main underlying verification technique of the \KeY\ system is \emph{symbolic execution}, a process during which  code fragments appearing in formulas are gradually transformed into \emph{state updates} -- syntactic representation of changes caused to the state of program variables by execution of the code. Symbolic, instead of concrete values are used for the variables, hence the name of the process. After symbolic execution is completed, the resulting first-order predicate logic formulas are proved, either within \KeY\ itself, or are handed over to external \emph{satisfiability modulo theories (SMT) solvers}, which may offer better performance.

A successful attempt of proving a formula indicates correctness of the method under verification with respect to its specification formalized in a method contract. \KeY\ system supports specifications written in \emph{Java Modeling Language (JML)} \cite{Leavens06} or its own input format. Proof search may be done interactively inside the graphical user interface by applying calculus rules to a formula (a list of applicable rules is suggested by \KeY) or it may be performed automatically by using predefined proof search strategies.

The question of how to handle method calls during symbolic execution is in the focus of this thesis. Currently \KeY\ offers two possibilities. The first one is to substitute a method call with its concrete implementation, which is a simple but highly ineffective approach in a scenario where the method implementation is expected to be frequently modified. The second possibility is to substitute the method call with a declarative specification of its effects (a postcondition). For this to work, two things are necessary: the called method has to be successfully verified against its contract and its precondition has to hold in the state when it is called. Using the method contract provides a great level of abstraction from the implementation, but considering that any significant change to the method is most likely to affect its behavior and would require an adaptation of its contract, the proof of the callee would have to be redone from scratch.

An idea how to further increase the degree of independence of a proof from specifications of called methods was proposed by H{\"a}hnle, Sch{\"a}fer and Bubel \cite{Haehnle13} under the name of \emph{abstract method calls}. According to this concept, a method call is substituted by an \emph{abstract} version of its contract that contains non-specific pre- and postconditions. Symbolic execution continues until only first-order logic proof goals remain, after which the actual definitions of pre- and postconditions are used to finish the proof. This way its significant part, mainly the symbolic execution of a program being verified, remains independent of the specifics of called methods and may be reused if a proof has to be repeated to accommodate modifications in method contracts it builds on. When complex source code is verified, such as programs containing loops and requiring user interaction for successful verification, the possibility to reuse parts of the proof and save user interaction effort becomes extremely valuable and improves the modularity and the automation of verification.

\section{Contributions}
A major goal of this work is to increase the level of abstraction from method calls during symbolic execution of a program and to seamlessly integrate the concept of abstract method calls into the workflow of program verification. It is carried out in context of the \KeY\ system and the JML. The main contributions of our work are the following:
\begin{itemize}
\item We extend the set of specification forms that may be defined abstractly with abstract locations sets and abstract class invariants. We provide supporting formal definitions and extend the JML to cover abstract specifications.
\item We adapt the JavaDL logic to correctly handle abstract method calls 
\item We adapt verification workflow of the \KeY\ system to facilitate automatic reuse of proofs by clearly separating and caching reusable proof parts. Necessary changes are made to the graphical user interface and to the proof search strategies. 
\end{itemize}

This thesis is structured as follows. Chapter \ref{ch:background} provides the background on the JML and the Java Card DL to the extent relevant for the introduction of the Abstract Method Call approach. In Chapter \ref{ch:amc} we give the detailed description of this approach, introduce formal definitions of abstract invariants and extended abstract contracts and describe changes to the JML. Chapter \ref{ch:implementation} focuses on the implementation of Abstract Method Call in the \KeY\ verification environment. In Chapter \ref{ch:evaluation} we report on the results of empirical evaluation of our approach. In Chapter \ref{ch:relwork} we discuss the research progress in the related area and in Chapter \ref{ch:conclusion} we conclude the thesis by summarizing the results and by setting up directions for future work.

\chapter{Background}\label{ch:background}
\section{Java Modeling Language}
The \emph{Java Modeling Language} is a language designed for writing specifications about Java programs. It combines the ideas of the Eiffel language \cite{Meyer98} and Larch language family \cite{Guttag93}, thus belonging to the class of specification languages following the design by contract concept. It has been developed since 1998 \cite{Leavens98} by a group of researches led by Leavens and has since then established itself as the de-facto standard in the area of design by contract style specifications for Java. 

In this section we provide a brief introduction of the JML features relevant for this thesis. 

\subsection{JML in source code}
JML specifications for a program may be provided in a separate file, but a more common style is to integrate them into the Java code as annotations. A \emph{JML annotation} is a standard Java comment, which is ignored by a Java compiler, but detected by JML tools thanks to an additional marking with an at-symbol (\texttt{@}). Both single line comments ( \texttt{//@ \invariant\ <boolean JML expression >}) and C-style comments (\texttt{/*@ \invariant\ <boolean JML expression> @*/)} are accepted. Such integration facilitates the reading of specifications and thanks to the rules regulating where the specification elements have to be inserted, clearly indicates the connections between them and the corresponding class elements.

\subsection{JML expressions}
\emph{JML expressions} are the basic building blocks of JML annotations used to write assertion predicates appearing directly after keywords like \invariant, \requires\ and \ensures\ and describing state properties. The structure of a JML expression can be shortly described as follows:
\begin{center}
Java expression -- side-effects + first-order logic + \old, \result, ...
\end{center}
The syntax of JML expressions builds on that of the Java expressions to enhance its practicability and to make specification writing more convenient for programmers without special knowledge of formal methods. However, some restrictions and extensions are necessary to bring the Java language up to a required level of expressiveness. On the side of restrictions, JML prohibits use of any Java expressions with side-effects, like \texttt{x = y + 10}, \texttt{i++} or a call to a state changing method, because a specification should have a purely logical meaning. 
On the other hand Java syntax is extended with some features of first-order logic:
\begin{itemize}
	\item logical connectives \keyword{==>} and \keyword{<==>} 
	\item quantified expressions \texttt{(\jforall\ T x; b1; b2)}, with a meaning that for all instances \texttt{x} of type \texttt{T} for which the expression b1 holds, expression b2 must hold too and \texttt{(\jexists\ T x; b1; b2)}, with a meaning that there exists an instance \texttt{x} of type \texttt{T} for which both expressions b1and b2 hold.\end{itemize}

And, last but not least, there are special keywords:
\begin{itemize}
	\item \old\texttt{(e)} is used in postconditions to refer to the value of an expression \texttt{e} right before a method execution
	\item \result\ is used in postconditions to refer to the return value of the method 
	\item \fresh\texttt{(e)} is a boolean expression that equals true, if an object to which \texttt{e} refers did not exist before a method execution 
\end{itemize}

\subsection{Method contracts and class invariants}
The JML focuses on the formalization of the intended methods' behavior of the class that is being verified. For this reason \emph{method contracts} are regarded as the main feature of the JML and moreover, they are of the highest interest for our work.

\begin{lstjavajml}[numbers=left, caption=Implementation of the class \java{Person} annotated with the JML, captionpos=b, frame = tb, basicstyle=\small, numberstyle=\tiny]
public class Person {
    private /*@ spec_public @*/ String name;
    private /*@ spec_public @*/ int weight;
    
    //@ public invariant !name.equals("") && weight >= 0; 
    
    /*@ public normal_behavior
       @ requires kgs >= 0;
       @ ensures weight == \old(weight) + kgs;
       @ assignable weight;
       @ also 
       @ public exceptional_behavior
       @ requires kgs < 0;
       @ signals (IllegalArgumentException) weight == \old(weight);
       @*/
    public void addKgs(int kgs) throws IllegalArgumentException {
        if (kgs >= 0) { weight += kgs;} 
        else { throw new IllegalArgumentException();}
    }
    
    //@ normal_behavior
    //@ ensures \result == weight;
    public /*@ pure @*/ int getWeight() {
        return weight;
    }
    
    public Person(String n) { ... }
}
\end{lstjavajml}

Let us consider an example taken from \cite{Leavens06Ex} and concentrate on the formal specification of the method  \texttt{addKgs} (lines 7 to 15). As we see, a method can have several contracts, separated by the keyword \also. A contract, as well as a class invariant, is preceded by a \emph{visibility modifier}, in this case it is \public. The JML inherits the visibility system of Java: a specification may not address a field or a method of lower visibility level than itself and a method contract may not be of higher visibility level than the method it specifies. To make an exception to these rules the keyword \specpublic\ is used (lines 2 and 3). It makes the fields \texttt{name} and \texttt{weight} public for specifications without affecting their Java visibility. 

The keywords \norm\ and \excep\ declare two different types of method contracts: the first one does not allow that an uncaught exception is thrown during the method execution, while the second one demands it. A precondition of the method can be declared with the keyword \requires\ and a postcondition (if we specify the normal behavior of the method) can be declared with the keyword \ensures. Their meaning is that if the precondition holds in the pre-state, then the method should terminate normally and the postcondition should hold in a post-state. Several requires clauses are joined by conjunction (same applies to postconditions). If a precondition misses, it means there are no obligations on the caller, which is equivalent to writing \requires\ \keyword{true} (missing postcondition implies \ensures\ \keyword{true}). The postcondition on line 9 tells, that after the normal execution of the method, the value of field \texttt{weight} (in the post-state) equals the value of \texttt{weight} $+$ \texttt{kgs} in the pre-state. In a contract for exceptional behavior the postcondition is declared by the keyword \signals\ followed by the type of an exception to be thrown and by the boolean JML exception specifying the post-state. Alternatively, the keyword \signalso\ may be used followed by the list of accepted exceptions only.

The \emph{assignable clause} specifies frame conditions, namely what locations may be assigned by the method. The keyword \assignable\ is followed by a list of so called \emph{store-ref} expressions or one of the keywords \everything, \nothing. Several assignable clauses are interpreted as a union and an absent assignable clause is equivalent to \assignable\ \everything. An assignable clause does not prohibit allocation of new objects or assignments to local variables of the method.

The \emph{class invariants} express consistent properties of a class object, that are established by the constructor and that must be preserved through object's lifetime. They must be preserved by method calls, that is if invariants hold before the method execution, they have to hold afterwards. To establish this requirement, invariants are implicitly included in pre- and postconditions of every public method and in postconditions of constructors. The invariant in line 9 states that the value of the field \texttt{name} never equals an empty string and that the value of the field \texttt{weight} is never negative. It would have been expected to include the condition \texttt{name != null} too, but actually it is a default requirement for any field in the current version of the JML (it may be changed by annotating a field with the keyword \nullable). The modifier \pure\ appearing in line 26 tells us, that the method \texttt{getWeight} has no side-effects and may therefore be called to in JML expressions.

Some other clauses, not appearing in this example but available in the JML are \emph{diverge clause} and \emph{accessible clause}. \diverge\ \texttt{b} states that the method is allowed to diverge, if the expression \texttt{b} holds in a pre-state and \accessible\ <store-ref list> defines locations that may be read by the method.

\subsection{Further features of the JML}
Loops might be annotated with \emph{loop invariants}, which are afterwards used to show the correctness of a postcondition. The keyword \loopinv\ is used to specify what stays invariant (should be valid in a state right before the first iteration and after an arbitrary number of loop iterations), the keyword \decreases\ indicates a variant integer expression which is strictly decreased by the loop body and which has to be $>= 0$ after every loop iteration to prove termination of the loop and the keyword \assignable\ states what locations may be modified by the loop body. The loop invariant annotations appear in the method body right before the loop itself. Other features of JML are data abstraction (\emph{model fields} and \emph{represents clauses}), \emph{assertion statements}, \emph{model programs}, etc., but they are not in focus of this thesis and therefore, are not explained in this section.

\section{Java Card Dynamic Logic}
For the purposes of design by contract verification we are in need of a logic, that could express a statement of the following kind: "if the precondition $P$ holds, then the program $p$ terminates and afterwards the postcondition $E$ holds". The suitable logic was proposed by \cite{Harel00} under the name \emph{dynamic logic (DL)}. It is an extension of typed first-order logic with two modality operators: \dlbox{\cdot} called \emph{box modality} and \dia{\cdot} called \emph{diamond modality}. A DL formula \dia{p}E with a program $p$ and a postcondition $E$ is used to formalize \emph{total correctness}, meaning that the formula is valid, if the program terminates and the postcondition holds afterwards. Alternatively, with the box modality one can express \emph{partial correctness}: if the program terminates, then the postcondition must hold, but non-termination is also allowed. The dynamic logic used in the \KeY\ system is tailored to the subset of Java programming language called \emph{Java Card} \cite{Beckert01} and in this section we present its syntax and semantics  in the extent relevant for this thesis. At the moment of writing we consider \cite{Keybook07} and \cite{Weiss11} to be the best sources for the full account of the \emph{Java Card DL}, which have also served as the basis for this section

\subsection{Syntax}
\begin{definition}[Signature]
\label{Signature}
A \java{Java Card DL} signature $\Sig$ is a tupel (\Types, $\subtype$, \Fun, \Pred, \Vars, \PVars, \sigfunc, \Prg), consisting of 
\begin{itemize}[noitemsep]
	\item a set \Types\ of type symbols (or simply \emph{types})
	\item a subtype relation $\subtype\ \subseteq \Types \times \Types$
	\item a set \Fun\ of rigid function symbols
	\item a set \Pred\ of rigid predicate symbols
	\item a set \Vars\ of rigid logical variables symbols
	\item a set \PVars\ of non-rigid program variables symbols
	\item a typing function $\sigfunc : \Fun \cup \Vars \cup \PVars \rightarrow \Types^{+}$. We write $\tau\ v$ instead of $\sigfunc(v) = \tau$, $p(\tau_1, ... , \tau_n)$ instead of $\sigfunc(p) = (\tau_1, ... , \tau_n)$ and $\tau\ f(\tau_1, ... , \tau_n)$ instead of $\sigfunc(f) = (\tau_1, ... , \tau_n, \tau)$ to denote the signatures of variable, predicate and function symbols.
	\item a program \Prg\ -- any valid set of \java{Java Card} classes and interfaces.
\end{itemize}
\end{definition}

\begin{figure}
\centering
\includegraphics[height=6cm]{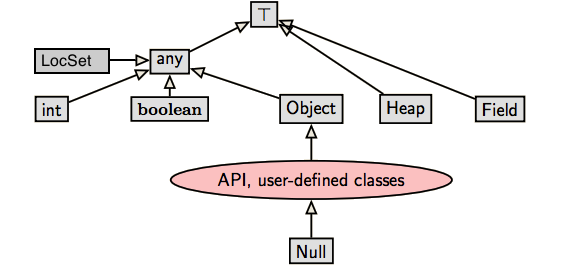}
\caption{Type Hierarchy for \java{Java Card DL}}
\label{types}
\end{figure}

The set \Types\ gathers all the types that may be used inside \java{Java Card DL} formulas. This includes basic types like \emph{boolean} and \emph{int}, but also all built-in API reference types (like \java{java.lang.Throwable}, \java{java.lang.Exception}, \java{java.util.AbstractCollection} and etc.) and any user-defined reference types. To reflect the concept of class inheritance in \java{Java Card} programs, we introduce a subtype relation $\subtype\ \subseteq \Types \times \Types$, which orders the elements of the set \Types. For example, classes \java{Vehicle} and \java{Car} of program \Prg\ both appear in the set \Types\ under the same names, and if the class \java{Car} inherits from the class \java{Vehicle}, this relationship will be captured by a tuple $(\mathit{Car}, \mathit{Vehicle}) \in \subtype$. Further basic types present in any \java{Java Card DL} signature are hierarchically depicted in  Figure \ref{types} \footnotemark.

\footnotetext{Source: \emph{Dynamic Logic} slides from \emph{Formale Grundlagen der Informatik 3}, TU Darmstadt, WS 2013/2014}

In the syntax of \java{Java Card DL} we strictly differentiate between \emph{rigid} symbols, whose interpretation is the same in all program states and \emph{non-rigid} (flexible) symbols, whose interpretation is dependent on the state. So for example, all typical elements of the predicates set ($<$, $<=$, etc.) and of the functions set ($\mathit{TRUE}$, $0$, $1$, $+$, $*$, etc.) are rigid. In contrast to predicate and function symbols, we distinguish two disjoint sets of variables \Vars\ and \PVars\ , such that $\Vars\ \cap \PVars\ = \emptyset$. Elements of the set \Vars\ are logical variables declared in first-order logic quantifiers and cannot be used in modalities. Elements of the set \PVars\ are program variables, whose semantics can be changed by the program and that can appear both in modalities and first-order logic parts of the \java{Java Card DL} formulas. Consider an example:
\[\exists \mathit{int}\; i; \dlbox{\mathtt{x} = 1;}(\mathtt{x} \keyeq i).\]
In this formula $\mathit{int}\ i \in \Vars$, $\mathit{int}\ \mathtt{x} \in \PVars$ and $\mathit{boolean}\ \keyeq(int, int) \in \Pred$.
\begin{note}
We require that for any Java field \java{f} of \Prg\ there is a constant symbol $f\in \Fun$ of type $\mathit{Field}$.\end{note}

\begin{note}
We model dynamic memory with a global program variable \java{heap} $\in \PVars\ $ of the type \mbox{$Heap \in \Types$}. We use the function symbol any select(Heap, Object, Field) to read from the heap at a specified position, the function symbol Heap store(Heap, Object, Field, Any) to write in the heap at a specified position, the function symbol Heap create(Heap, Object) to create an object on the heap and the function symbol Heap anon(Heap, LocSet, Heap) to anonymise the heap.
\end{note}

\begin{definition}[Term]
\label{Term}
A term of type $\tau \in \Types$ is either 
\begin{itemize}[noitemsep]
	\item a variable $v \in \Vars \cup \PVars$ of the type $\tau$
	\item a function $f (t_1,..., t_n)$ with the result type $\tau$ and $t_1,..., t_n$ -- terms of correct types with accordance to the signature of $f$.
	\item an \emph{updated term} $\{u\}t$ for an update $u$ (Definition \ref{Update}) and a term $t$ of type $\tau$.
\end{itemize}
\end{definition}

\begin{note}
We use a separate type $\mathit{LocSet}$ to refer to a set of pairs $(o, f)$, where $o$ is a term of type $\mathit{Object}$ and $f$ is a constant symbol of type $\mathit{Field}$. We add function symbols $\dotcup$, $\dotcap$ and $\dotsetminus$ and predicate symbols $\dotin$, $\dotsubseteq$ defined on the type $\mathit{LocSet}$.
\end{note}

The update concept is one of the most principal in \java{Java Card DL}. It serves as a mean to keep track of state changes occurring during symbolic execution of the program appearing in a formula (Definition \ref{Formula}). Its main difference from the more common approach of substitutions is that it allows to simplify and resolve conflicts of state changes at an early stage, postponing the application of substitutions to the variables till only after the symbolic execution of the whole program is complete.

\begin{definition}[Update]
\label{Update}
An update is either:
\begin{itemize}[noitemsep]
	\item a simple update $\mathit{pv} := t$ for a program variable $\mathit{pv}$ and a type consistent term $t$
	\item a parallel update $u_1 || u_2$ for updates $u_1$ and $u_2$
\end{itemize}
\end{definition}

For example, a \java{Java Card DL} formula $\dia{x = 1; x = 2}(x > 0)$ is transformed into $\{x := 1\}\dia{x = 2}(x > 0)$, then into $\{x := 1 || x := 2\}\dia{}(x > 0)$, where a parallel update is simplified into $\{x := 2\}\dia{}(x > 0)$ and then applied to a variable $2 >0$

\begin{definition}[Formula]
\label{Formula}
A \java{Java Card DL} formula is either 
\begin{itemize}[noitemsep]
	\item TRUE, FALSE
	\item a predicate $p (t_1,..., t_n)$ with $t_1,..., t_n$ -- terms of correct types with accordance to the signature of $p$
	\item $\neg \phi$, $\phi \wedge \psi$, $\phi \vee \psi$, $\phi \rightarrow \psi$, $\phi \leftrightarrow \psi$, $\forall \tau\ x\; \phi$ and $\exists \tau\ x\; \phi$ for any \java{Java Card DL}\ formulas $\phi$, $\psi$, any variable $x \in \Vars$ and any type $\tau \in \Types$.
	\item $\dlbox{p} \phi$ and $\dia{p} \phi$ for any program $p$ and a \java{Java Card DL}\ formula $\phi$
	\item an updated formula $\{u\}\phi$ for any update $u$ and a \java{Java Card DL}\ formula $\phi$
\end{itemize}
\end{definition}

\subsection{Semantics}
As usual in dynamic logic, we define the semantics in terms of a \emph{Kripke Structure}:

\begin{definition}[Kripke Structure]
\label{Kripke Structure}
A \java{Java Card DL} Kripke structure is a tuple $\Kripke = (\Domain, \delta, \Interp, \State, \rho)$ consisting of 
\begin{itemize}[noitemsep]
	\item a domain \Domain\ and a typing function $\delta: \Domain \rightarrow \Types$
	\item an interpretation function \Interp\ for predicate and function symbols
	\item a set of \emph{states} \State\ consisting of functions $s \in \State$ mapping any program variable $\mathit{pv}\in \PVars$ to a value $d \in \Domain$
	\item a transition function $\rho$, such that for any program $p$ and any two states $s_1, s_2 \in \State$ it is defined as $\rho(s_1, p) =  s_2$ iff the program $p$ started in the state $s_1$ terminates and its final state is $s_2$, otherwise it is undefined.
\end{itemize}
\end{definition}
The above definition of a transition function implies, that any program is deterministic, but does not necessarily terminate.
We omit the exact definitions of the domain \Domain\ and the interpretation function \interp\ which can be found in their full extent in \cite{Weiss11}.

\subsection{Calculus}

Similar to the first-order logic we show the validity of a \java{Java Card DL} formula in terms of the sequent calculus. The sequent $\sequent{}{}$ is said to be valid, if for any state $s \in \State$ holds $s \models \sequent{}{}$. The validity proof of a sequent $\sequent{}{}$ is constructed in form of a \emph{proof tree} of \emph{calculus rules}. We use the classical notation notation to define the calculus rules:
\[\seqRule{ruleName}{p_1 \dots p_n}{c}\] 
which means that the \emph{conclusion} sequent $c$ is logically valid if all premisses $p_1 \dots p_n$ (also sequents) are logically valid. Alternatively we define \emph{rewrite rules} in form $\mathit{left} \leadsto \mathit{right}$ which means that an occurrence of the schematic term $\mathit{left}$ in any formula can be replaced with the term $\mathit{right}$, for example $t \doteq t \leadsto \mathit{TRUE}$.

The calculus of \java{Java Card DL} consists of hundreds of rules that may be divided into two groups: first-order logic calculus rules and rules for symbolic execution of the programs appearing in modalities. The latter handles the control flow of the program, simplification of complex expressions and modeling of state changes. The symbolic execution calculus rules are applied to the first active statement of the program. For example, assignments to local variables and object fields are converted into equivalent updates:
\[\seqRule{assignLocal}{\sequent{}{\{\mathtt{x} := t\}\dia{\omega}\phi}}{\sequent{}{\dia{\mathtt{x} = t; \omega}\phi}}\;  \seqRule{assignField}{\sequent{}{\{\java{heap} := \mathit{store}(\java{heap}, o, \java{f}, t)\}\dia{\omega}\phi}}{\sequent{}{\dia{o.\java{f} = t; \omega}\phi}}\]
Complex expressions as \java{i = x++;} have to be transformed into several simple sub-expressions \java{int temp; temp = x + 1; i = k} using temporal variables. Update applications may be carried out on variables, functions, predicates and first-order logic formulas, for example:
\begin{align*}
\{\mathtt{x}:=t\}\mathtt{y} \leadsto \mathtt{y} &&  \{\mathtt{x}:=t\}\mathtt{x} \leadsto t
\end{align*}
\[ \{\mathtt{x}:=t\}f(t_1, \dots, t_n) \leadsto f(\{\mathtt{x}:=t\}t_1, \dots, \{\mathtt{x}:=t\}t_n)\]
where $\mathtt{x}, \mathtt{y}$ are program variables, $\mathtt{x} \not= \mathtt{y}$, $f$ is a rigid function symbol and $t_1, \dots, t_n$ are terms.

The handling of method calls is of special interest for this thesis. As an alternative to substituting the method call inside the calling program with the body of called method its contract can be used. A simplified version of a \java{useMethodContract} rule is given in  Definition \ref{useMethodContract}. It can be applied only in the context where partial correctness is to be proven and thus, does not place any demands on the termination of the called method. Furthermore, we omit the requirements on the well-formedness of the heap and the reachability of input and output variables, as being not introduced in this thesis and not relevant to our topic. The definition of the \java{useMethodContract} rule in its full extent can be found in \cite[p. 147]{Weiss11}
\begin{definition}[Rule \java{useMethodContract}] \mbox{}\\
\begin{center}
$\seqRule{}{\sequent{} {\{u\}\{w\}(\pre)}\\ 
\sequent{}{\{u\}\{w\}\{\hPre := \heap\}\{v\}(\post \rightarrow \dlbox{\java{r} = \res; \omega}\phi}}
{\sequent{}{\{u\}\dlbox{\ \java{r} = o.\java{m}(a_1, \dots a_n);\ \omega}\phi}}$
\end{center}
where:

\begin{itemize}[noitemsep]
	\item $o$ is the receiver of the method \java{m} and $a_1, \dots a_n$ are its actual parameters 
	\item the \java{Java Card DL} formulas $\pre$ and $\post$ are the pre- and postcondition of method \java{m}'s contract
	\item the term $\mymod$ of the type $\mathit{LocSet}$ is the modifies clause of method \java{m}'s contract
	\item the program variable $\res$ represents the result value of the method \java{m} in the formulas $\pre$ and $\post$
	\item $w = (\self := o\ ||\ \java{p}_1 := a_1\ ||\ \dots \ ||\ \java{p}_n := a_n)$, where \self\ and $\java{p}_1 \dots \java{p}_n$ are program variables representing the receiver object of the method and its parameters in the formulas $\pre$ and $\post$ and the term $\mymod$.
	\item $v = (\java{heap} := \mathit{anon}(\java{heap}, \mymod, h')\ ||\ \res : = a)$, where $h'$ and $a$ are fresh constant symbols.
\end{itemize}
\label{useMethodContract}
\end{definition}

The first premiss of this rule formalizes the requirement, that in the state at invocation time, the precondition of the contract should hold.
In a different branch, which corresponds to the second premiss, the postcondition is used as an assumption to prove the validity of the rest of the formula at a state after the call. The update $w$ replaces all the formal variables used in the contract with their actual values at the time of the call (the object and the passed arguments). The update $v$ used in the second premiss is called an \emph{anonymising update}. It formalizes an idea, that without performing the explicit symbolic execution of the body of method \java{m}, we cannot know the values stored on the heap, as well as the result value, in the state after the call, so they are updated with \emph{unknown} values with help of \emph{fresh} symbols. In contrast to $\res$ variable, the anonymisation of heap is handled in a special way: only locations mentioned in $\mymod$ clause have to be anonymised -- other locations are specified as non-modifiable and therefore their values stay the same as in the state before the call. The semantics of a subsequent select operation on such an anonymised heap is reflected by the following simplified (the aspects of object creation are ignored) rewrite rule, which full version can be found in \cite[p. 94]{Weiss11}):
\begin{align*}
\mathit{select(anon(h, s, h'), o, f)} \leadsto \mathit{if\ (o, f)} &\in s\\
&\mathit{then\ (select(h', o, f))\ else\ (select(h, o, f))} 
\end{align*}
where $h$ is the "original" heap and $h'$ is the "anonymous" heap and which states, that for any location not in $s$ \emph{select} returns the value from the original heap.

Note, that local variables of the caller also stay unchanged, because they are not visible to the method \java{m}.

A similar idea is used for loops appearing inside the program. They pose exceptional difficulty to symbolic execution, mainly because expanding the loop body only makes sense if the bound on the number of loop iterations is literally known in advance, which is seldom the case. Instead, the body of the loop is substituted with the loop invariant. In Definition \ref{loopInvariant} we present a simplified version of a \java{loopInvariant} rule. It is applicable in the context of partial correctness and does not establish the termination of the loop. Furthermore, for the sake of simplicity it assumes that the loop body does not contain \textbf{\java{return}}, \textbf{\java{break}} or \textbf{\java{continue}} and does not throw exceptions. As for the \java{useMethodContract} rule, we omit the requirements on the well-formedness of the heap and the reachability of input and output variables. The full version can be found in \cite[p. 108]{Weiss11}
\begin{definition}[Rule \java{loopInvariant}] \mbox{}\\
\begin{center}
$\seqRule{}{\sequent{}{\{u\}(\mathit{inv})}\\ 
\sequent{}{\{u\}\{\pre\}\{v\}(\mathit{inv} \wedge g \keyeq TRUE \rightarrow \dlbox{p}(\mathit{inv} \wedge \myframe)}\\ 
\sequent{}{\{u\}\{\pre\}\{v\}(\mathit{inv} \wedge g \keyeq FALSE \rightarrow \dlbox{\omega}\phi}}
{\sequent{}{\{u\}\dlbox{\java{while}(g)p;\ \omega}\phi}}$
\end{center}
where:

\begin{itemize}[noitemsep]
	\item a \java{Java Card DL} formula $\mathit{inv}$ is a loop invariant
	\item a term $\mymod$ of the type $\mathit{LocSet}$ is a modifies clause of the loop
	\item $\java{b}_1, ..., \java{b}_n \in \PVars$ are the program variables that can potentially be modified by the loop body
	\item $v = (\heap:= anon(\heap, \mymod, h)\ ||\ \java{b}_1 := b'_1\ ||\ ...\ ||\ \java{b}_n := b'_n)$, where $\java{h}, b'_1, \dots, b'_n$ are fresh constant symbols.
	\item $\pre = (\hPre := \heap\ ||\ \java{b}_1^{\pre} := \java{b}_1\ ||\ \dots\ ||\ \java{b}_n^{\pre} := \java{b}_n)$, where $\hPre, \java{b}_1^{\pre}, \dots, \java{b}_n^{\pre}$ are fresh program variables.
	\item a \java{Java Card DL} formula 
		\begin{multline*}
			\myframe :=\forall \Object\ o; \forall \Field\ f; ((o, f) \dotin \{\pre'\}\mymod\\ \vee select(\heap, o, f) \doteq select(\hPre, o, f))
		\end{multline*}
	\item $\pre' = ((\heap := \hPre\ ||\ \java{b}_1 := \java{b}_1^{\pre}\ ||\ \dots \ ||\ \java{b}_n := \java{b}_n^{\pre})$
\end{itemize}
\label{loopInvariant}
\end{definition}

The first premiss of the rule requires, that the invariant holds in the state before the first iteration. The second premiss is close to an induction step: assuming that the invariant holds after an arbitrary number of iteration, it has to be proven, that it holds after another one ($p$ -- the body of the loop -- has to be analyzed once). To reflect that the state has changed after an arbitrary number of iteration we again use an \emph{anonymising} update $v$. In contrast to method contract rule, it also includes potentially modifiable (appearing on the left side of the assignment statements inside the loop body) local variables. A separate task is to prove that the frame condition (a statement, that only locations appearing in the modifiable clause may be changed by the loop body) holds. The update $\pre$ buffers the values of the state before the first iteration of the loop, and the update $\pre'$ uses them to inverse the anonymisation for $\mymod$ term only. The last premiss formalizes the exit out of the loop. Assuming that the invariant holds after all iterations of the loop, the rest of the formula has to be validated.

\chapter{Abstract Method Call} \label{ch:amc}

\newcommand{\vars}{\ensuremath{\mathit{PVars}}}
\newcommand{\plhs}{\ensuremath{\mathit{Plhs}}}
\newcommand{\defs}{\ensuremath{\mathit{Defs}}}
\newcommand{\theap}{\ensuremath{\java{heap}}}

\section{Abstract contracts at a glance}
In the introduction to this thesis we have spoken of modular verification as the motivation for our work. More exactly we have identified as our goal the improvement of the reusability of proofs that rely on method contracts. A typical scenario in which the problem of such reuse arises, is illustrated by our running example, which we introduce now.\\

\begin{lstjavajml}[numbers = left, caption=Implementation of the class \java{StudentRecord}, captionpos=b, frame = tb, label = student, basicstyle=\small]
class StudentRecord {
	// exam result
	int exam;
	
	// achieved bonus
	int bonus;
	
	// minimum grade necessary to pass the exam
	int passingGrade;
	
	// completed labs
	boolean[] labs;
	
	//@ public invariant exam >= 0 && bonus >= 0 && passingGrade >= 0;
	//@ public invariant labs.length == 10;

	/*@... @*/
	int computeGrade() { ... }
	
	/*@
	@ public normal_behavior
	@ ensures \result ==> exam + bonus >= passingGrade;
	@ ensures \result ==> (\forall int x; 0 <= x && x < 10; labs[x]);
	@ assignable \nothing;
	@*/
	boolean passed() {
		boolean enoughPoints = computeGrade() >= passingGrade;
		boolean allLabsDone = true;
		for (int i = 0; i < 10; i++) {
	 	   allLabsDone = allLabsDone && labs[i];
		}
		return enoughPoints && allLabsDone;
	}
}
\end{lstjavajml}

As an example scenario for advantageous proof reuse we want to consider the \java{StudentRecord} class shown in Listing \ref{student}. This class represents results of a student's participation in an arbitrary university course and provides functionality to determine, whether he or she has passed the course.

A course is considered to be successfully completed (method \java{passed()}), if all the labs were done \footnotemark and the performance at the exam was satisfactory. The latter can be improved by the bonus (\java{computeGrade}), for which two different practices have been established. One approach, presented in Listing \ref{version 1}, says that bonus is simply added to the exam result unconditionally, the other, shown in Listing \ref{version 2}, places a restriction, that bonus can only be used to improve the exam grade, if the exam itself was passed, i.e., the obtained grade is \java{>= 4}. Thus, we get two versions of the \java{StudentRecord} class, which only differ in the implementation of the \java{computeGrade()} method.
\begin{figure}[h]
\begin{mdframed}[leftline =false, rightline=false]
\begin{minipage}{.40\textwidth}
\begin{lstjavajml}[caption=Version 1, captionpos=b, breaklines=true, label = {version 1}, basicstyle=\small]
class StudentRecord {
    {...} // Fields as in Listing 3.1
    /*@
    @ public normal_behavior
    @ requires bonus >= 0;
    @ ensures \result 
    	== exam + bonus;
    @ assignable \nothing;
    @*/
	
    int computeGrade(){
        return exam + bonus;
    }
    
    /*@ ... @*/
    boolean passed() {...}
}
\end{lstjavajml}
\end{minipage}\hfill
\begin{minipage}{.50\textwidth}
\begin{lstjavajml}[caption=Version 2, captionpos=b, breaklines=true, label = {version 2}, basicstyle=\small]
class StudentRecord {
    {...} // Fields as in Listing 3.1
    /*@
    @ public normal_behavior
    @ requires bonus >= 0;
    @ ensures (exam >= passingGrade) 
    	==> \result == exam + bonus;
    @ ensures (exam < passingGrade) 
    	==> \result == exam;
    @ assignable \nothing;
    @*/

    int computeGrade(){
        if (exam >= passingGrade) {
            	return exam + bonus;
        } else {
	   		return exam;
        }}
        
    /*@ ... @*/
    boolean passed() {...}
}
\end{lstjavajml}
\end{minipage}
\end{mdframed}
\caption{Two alternative implementations of the method \java{computeGrade()}}
\end{figure}

\footnotetext{Please note, that for our example the length of the array \texttt{labs} is irrelevant. In the proof for the \texttt{passed()} method the loop is handled by the \texttt{loopInvariant} rule and, as a short study has sown, using fixed length has no significant influence over the structure of the proof. The loop annotation can be found in Appendix.}
As we can observe, method \java{passed()} has a call to method \java{computeGrade()} inside its body. An important characteristic of this example is that not only the implementation of method \java{computeGrade()} is different in the two presented versions of the \java{StudentRecord} class, but also its specification. If only the first were the case, the existing \java{useMethodContract}, as introduced in Section 2.2.3 (Definition \ref{useMethodContract}), would provide the desired level of abstraction from the implementation, but in the given scenario it fails. Let us assume that upon arrival to this call during symbolic execution of method \java{passed()}, the call is handled by the \java{useMethodContract} rule. This makes any subsequent proof steps potentially specific to the concrete information provided by the \java{computeGrade()} contract, for example that the \java{computeGrade()} method from Version 1 (Listing \ref{version 1}) does not modify any location. Thus, the rest of the proof of the correctness of method \java{passed()} from Version 1 could not be reused to prove the correctness of the same method \java{passed()} from Version 2 (Listing \ref{version 2}), even though their implementations are identical. To further enhance the problem, we point out, that the difference in modifies clauses of the \java{computeGrade()} contracts propagates the change in the modifies clause of the \java{passed} contract.

A solution to this problem was proposed by H{\"a}hnle, Sch{\"a}fer and Bubel \cite{Haehnle13} under the name of \emph{abstract method calls}. Its main idea is to postpone any first-order logic reasoning based directly on the information from the used method contract to the latest possible stage, thus enlarging the reusable part of the proof. To achieve this a proof operates on placeholders standing for pre- and postconditions of used contracts until at least the end of symbolic execution, instead of using their actual content. A contract defined with placeholders received the name \emph{abstract method contract}, while its components are referred to as \emph{abstract precondition} and \emph{abstract postcondition} or \emph{abstract requires clause} and \emph{abstract ensures clause}.

However, at the proposed in \cite{Haehnle13} form an abstract contract is subject to a serious limitation, namely the fact that the assignable clause cannot be abstractly defined. As it was illustrated by our example, the modifies clauses are not less likely to be affected by change of method implementation the pre- and postconditions are. The recent changes in the modeling of heap \cite{Weiss11} in \java{Java Card DL} has made it possible to lift this restriction. Moreover, in the verification context supporting class invariants (such as \KeY\ system), which are implicitly added to pre- and postconditions of all class methods, it is essential to also bring them to an abstract form.

In the following two sections we define an abstract method contract in terms of JML and \java{Java Card DL}, addressing the named two issues by introducing an \emph{abstract assignable clause} to an abstract contract and completing it with an \emph{abstract class invariant}.

\section{Extending the JML Specification Language}
This section presents syntactical extensions introduced to JML to support the definition of abstract method contracts and abstract class invariants. Within the method contract our focus lies on \requires, \ensures\ and \assignable\ clauses, that is why for simplicity reasons we only address the definition of an abstract contract for specifying normal behavior.

\subsection{Abstract method contract} Listing \ref{template} presents a template with which an abstract method contract for the normal behavior can be specified in JML. This template contains two parts:
\begin{itemize}
	\item declaration of clauses (lines 2 -- 4)
	\item definition of clauses (lines 5 -- 7)
\end{itemize}
We consider the requires, ensures and assignable clauses of the contract to be completely \emph{desugared}, meaning that all of the specifications usually implicitly added to the contract by the JML syntactic sugar (like a clause \assignable\ \nothing\ for a method annotated with the keywords \pure) are explicitly formalized in the clauses themselves (except for class invariants).

\begin{center}
\begin{lstjavajml}[caption=Template for an abstract method contract, captionpos = b, label = template, numbers=left, basicstyle=\small]
    @ <visibility modifier> normal_behavior
    @ requires_abs R;
    @ ensures_abs E;
    @ assignable_abs A;
    @ def R = <boolean JML Expression>; 
    @ def E = <boolean JML Expression>; 
    @ def A = <store-ref list>;
\end{lstjavajml}
\end{center}

\textbf{Declaration of clauses}. In this part the keywords \rabs, \eabs\ and \aabs\ are used to declare \emph{placeholders} for requires, ensures and assignable clauses correspondingly. These placeholders are unique symbols, representing, in case of a requires and ensures clauses, a boolean JML expression and in case of an assignable clause -- a set of locations. Their semantics with respect to the correctness of a method remain unchanged from that of a concrete method contract. So, for instance, in the above form of an abstract contract, the placeholder P stays for the precondition of the method, the placeholder E stays for the postcondition of the method and the placeholder A stays for the list of assignable locations of the method. Moreover, as in a concrete method contract, several placeholders for requires clauses may be declared, which is interpreted exactly as several requires clauses appearing in one contract (the same holds for placeholders for ensures and assignable clauses). When an abstract contract is translated into the \java{Java Card DL}, the placeholders are interpreted as predicates or functions of the same name, with functions having the return type $\mathit{LocSet}$ (see Definition \ref{def:placeholder}).

\textbf{Definition of clauses}. The second part consists of an arbitrary number of \emph{definition clauses}. Those are used to provide concrete instantiations  to the placeholders declared in the first part. A definition clause starts with the keyword \defabs, followed by the placeholder symbol to be defined and its definition. Note that the definition must correspond to the "type" of the placeholder it defines. For a placeholder declared with the keyword \rabs\ or \eabs\ it is a boolean JML expression, identical to those appearing after \requires\ and \ensures\ keywords in a concrete contract. For a placeholder declared with the keyword \aabs\ it is a list of \emph{store-ref} expressions or one of the keywords \everything, \nothing. The definition clauses are translated into the \java{Java Card DL} as rewrite rules, which will be explained in further details in Section \ref{sec:amc}.

\subsection{Abstract invariant} The idea pioneered on method contracts can be without any obstacles transferred onto class invariants. 
As seen in Listing \ref{env}, a template for an abstract invariant also consists of two parts: placeholders are declared with the keyword \keyword{invariant\_abs} and defined with the keyword \defabs . As usually, several placeholders can be declared and defined, in which case they will be joined by conjunction.
\begin{lstjavajml}[caption=Template for an abstract invariant, captionpos = b, label = env, numbers=left]
    @ <visibility modifier> invariant_abs env;
    @ def env = <boolean JML Expression>; 
\end{lstjavajml}
In accordance to the usual semantics of class invariants, placeholders standing for invariants are added to pre- and postconditions of the method contract by conjunction, whenever this contract is used.

\subsection{Example}
On the example of the earlier introduced \java{StudentRecord} class, we show in Listing \ref{studentabs} how a class can be annotated with abstract contracts and invariants.\\

\begin{lstjavajml}[numbers = left, basicstyle=\small, label = studentabs, caption={Abstractly annotated \java{StudentRecord} class, Version 1}, captionpos=b, frame = tb]
class StudentRecord{
	{...} // Fields as in Listing 3.1
	
	//@ public invariant_abs env1; 
	//@ def env1 = exam >= 0 && bonus >= 0 && passingGrade >= 0;
	//@ public invariant_abs env2;
	//@ def env2 = labs.length == 10;

	/*@
	@ public normal_behavior
	@ requires_abs computeGradeR; 
	@ ensures_abs computeGradeE;
	@ assignable_abs computeGradeA;
	@ def computeGradeR = bonus >= 0;
	@ def computeGradeE = \result == exam + bonus;
	@ def computeGradeA = \nothing;
	@*/
	int computeGrade(){
		return exam + bonus;
	}

	/*@
	@ public normal_behavior
	@ requires_abs passedR;
	@ ensures_abs passedE1;
	@ ensures_abs passedE2;
	@ assignable_abs passedA;
	@ def passedR = true;
	@ def passedE1 = \result ==> exam + bonus >= passingGrade;
	@ def passedE2 = \result ==> (\forall int x; 0 <= x && x < 10; labs[x]);
	@ def passedA = \nothing;
	@*/
	boolean passed() {
		boolean enoughPoints = computeGrade() >= passingGrade;
		boolean allLabsDone = true;

		for (int i = 0; i < 10; i++) {
	 	   allLabsDone = allLabsDone && labs[i];
		}
		return enoughPoints && allLabsDone;
	}
}
\end{lstjavajml}

\section{Abstract method call in \texttt{Java Card DL}}\label{sec:amc}
In this section we give the \java{Java Card DL} representation of abstract JML specifications introduced earlier. Definitions in this section apply to contracts of the methods (not constructors), that are neither static nor void. Extending them to cover these alternatives is straightforward. 

\subsection{Placeholder}
\begin{definition}[Placeholder] \label{def:placeholder}
Let $\java{C} \in \Types$ be a \java{Java Card} class and let \java{m} be a method declared in this class with parameters types $\tau_1 \dots \tau_n \in \Types$ and a return type $\tau \in \Types$. A placeholder is an uninterpreted predicate or a function symbol of one of the following four types:
\begin{itemize}[noitemsep]
		\item Requires placeholder is a predicate $R(\mathit{Heap}, C, \tau_1 \dots \tau_n) \in \Pred$, which depends on the heap at invocation time, the receiver object and the parameters of the method.
		\item Ensures placeholder is a predicate $E(\mathit{Heap},\; \mathit{Heap}, C, \tau, \tau_1 \dots \tau_n) \in \Pred$, which depends on the heap at the method's final state (first argument), the heap at invocation time (second argument), the receiver object, the result value and the parameters of the method.
		\item Assignable placeholder is a function $A(\mathit{Heap}, C, \tau_1 \dots \tau_n) \in \Fun$ of type $\mathit{LocSet}$, which depends on the heap at invocation time, the receiver object and the parameters of the method.
		\item Invariant placeholder is a predicate $I(\mathit{Heap},\; C) \in \Pred$, which depends on the current heap and the object for which the invariant should hold (for example, the receiver object of the method).
\end{itemize}
\end{definition}
We define placeholders as functions or predicates of multiple parameters to avoid faulty simplifications performed over them by \java{Java Card DL} calculus rules. The list of parameters on which the placeholder depends is defined following the nature of the specification the placeholder stands for. In the concrete requires clause and modifies clause it is possible to refer to the current heap (which is the heap at invocation time), the receiver object or the parameters of the method. In the ensures clause in addition to the current heap (which is now the heap in the method's final state), the receiver object and the parameters, one can refer to the heap at pre-state or the result value. In the invariants only the program variables representing the current heap and the receiver object of the method are accessible.

The keyword with which the placeholder is declared defines its type. For example, the declaration \java{\eabs\ computeGradeE} (line 12) from the specification of method \java{computeGrade()} from Listing \ref{studentabs} is translated into the ensures placeholder 
\[\mathit{boolean}~ \mathit{computeGradeE}(\mathit{Heap},\; \mathit{Heap},\; \mathit{StudentRecord},\; \mathit{boolean})\]
where given the program variables $\theap$, $\hPre$, $\self$ and $\res$ are used to refer to the current heap, heap at pre-state, its receiver object and its return value respectively in the specifications of the method, the placeholder parameters would be instantiated as follows 
\[\mathit{computeGradeE}(\theap,\; \hPre,\; \self,\; \res)\]
when they appear in the clauses of the abstract method contract.

\subsection{Invariant}
Using the invariant placeholder, we can give definition to an abstract invariant.

\begin{definition}[Abstract invariant]\label{def:invariant}
Let $\java{C} \in \Types$ be a \java{Java Card} class. Let \heap\ be a global program variable used to refer to the current heap.
An abstract invariant of the class \java{C} is a tupel 
\[(I, \self, \mathit{def_I})\] that consists of
\begin{itemize}
\item a unique invariant placeholder $I(\mathit{Heap},\;C )$ for class \java{C};
\item a unique program variable $\self$ declared for this class invariant to refer to the object of the class.
\item a term $\mathit{def_I}$ of type boolean which specifies the semantics of
  $I$ and that can make use of the two variables $\theap$ and $\self$
\end{itemize}
\end{definition}

\subsection{Abstract method contract}
Further making use of other types of placeholders, we can build the abstract clauses for the abstract method contract:
\begin{definition}[Abstract clauses]\label{def:absclauses}
The abstract preconditions $\mathit{pre^a}$, the abstract postconditions $\mathit{post^a}$ and the abstract modifies clauses $\mathit{mod^a}$ are built according to the following grammar:
\begin{align*}
\mathit{pre^a} &::= R\; |\; I \wedge \mathit{pre^a} \;|\; \mathit{pre^a} \wedge \mathit{pre^a}\\
\mathit{post^a} &::= E \wedge I \;|\; E \wedge \mathit{post^a} \;|\; I \wedge \mathit{post^a}\\
\mathit{mod^a} &::= A\; |\; \mathit{mod^a} \cup \mathit{mod^a} 
\end{align*}
where $R$, $E$, $A$ and $I$ are atomic formulas with a requires, ensures, assignable or invariant placeholder as top level symbol.
\end{definition}

This definition shows what kind of restrictions we place on abstract clauses. An abstract precondition should contain at least one requires placeholder and optionally a number of invariant placeholders joined by conjunction. An abstract postcondition should contain at least one ensures placeholder and at least one invariant placeholder, also joined by conjunction. An abstract modifies clause contains at least one assignable placeholder. 

As follows from Definition \ref{def:placeholder} of placeholders, the abstract clauses are, in case of an abstract precondition and an abstract postcondition, the non-rigid terms of type boolean and, in case of an abstract modifies clause, the non-rigid terms of type $\mathit{LocSet}$. 
 
Now we can define an abstract method contract. This definition applies to contracts that do not allow for exceptional termination.
\begin{definition}[Abstract method contract]
Let $\java{C} \in \Types$ be a \java{Java Card} class and let \java{m} be a method declared in this class with parameters types $\tau_1 \dots \tau_n \in \Types$ and a return type $\tau \in \Types$. Let $\heap \in \PVars$ be a global program variable used to refer to the current heap.
An abstract method contract for a method \java{m} is a tupel 
\[(\vars,\; \plhs,\; \pre^a,\; \post^a_{norm},\; \mymod^a,\; \mathit{term},\; \defs)\] consisting of

\begin{itemize}[noitemsep]
	\item a set of unique program variables $\vars = \{\hPre,\; \self,\; \res,\; \exc,\; \java{a}_1, \dots, \java{a}_n\}$ declared for the method \java{m} to refer to the heap at pre-state, the receiver object, the return value, the exception thrown by the method and the method parameters in its specifications. 
	\item a set of placeholders $\plhs$
	\item an abstract precondition $\pre^a$
	\item a postcondition $\post^a_{norm} = \post^a \wedge (\exc \keyeq \mathit{null})$, where $\post^a$ is an abstract poscondition
	\item an abstract modifies clause $\mymod^a$ 
	\item a termination modifier $\mathit{term} = \{\mathit{partial}, \mathit{total}$
	\item a set $\defs$ of pairs $(P, \mathit{def}_P)$ with a \java{Java Card} DL term $\mathit{def}_p$ of type $\mathit{boolean}$ or $\mathit{LocSet}$ for each placeholder $P \in \plhs$. A term $\mathit{def}_p$ specifies semantics of $P$ and can make use of variables from the set $\vars$
\end{itemize}
\label{abscontract}
\end{definition}

The abstract clauses $\pre^a,\; \post^a,$ and $\mymod^a$ are constructed according to the Definition~\ref{def:absclauses} from placeholders belonging to the set $\plhs$. This set contains requires, ensures and assignable placeholders declared for the particular method \java{m} and the invariant placeholders declared for the class \java{C}. This set must contain at least one placeholder of each type. The parameters of the placeholders are instantiated with program variables from the set $\vars$ in the following manner:
\begin{itemize}	[noitemsep]
	\item Requires placeholder -- $R(\theap,\; \self,\; \java{a}_1, \dots, \java{a}_n)$
	\item Ensures placeholder -- $E(\theap,\; \hPre,\; \self,\; \res,\; \java{a}_1, \dots, \java{a}_n)$
	\item Assignable placeholder -- $A(\hPre,\; \self,\; \java{a}_1, \dots, \java{a}_n)$
	\item Invariant placeholder -- $I(\theap,\; \self)$
\end{itemize}

The program variable $\exc$ represents an exception thrown by the method, or $\mathit{null}$, if no exception is thrown.
A boolean term $\mathit{def}_p$ defining the semantics of a requires placeholder can make use of the program variables $\theap,\;  \self,\; \java{a}_1, \dots, \java{a}_n$. A boolean term $\mathit{def}_p$ defining the semantics of an ensures placeholder can additionally make use of the program variables $\hPre$ and $\res$. A term $\mathit{def}_p$ of type $\mathit{LocSet}$ defining the semantics of a modifies placeholder can feature the program variables $\hPre,\;  \self,\; \java{a}_1, \dots, \java{a}_n$. A boolean term $\mathit{def}_p$ defining the semantics of an invariant placeholder can make use of the program variables $\theap$ and  $\self$

\subsection{Example}
To illustrate the definition of the abstract method contract we provide a translation of the JML specification of the method \java{computeGrade()} (lines 9 -- 17) from the Listing \ref{studentabs} into \java{Java Card} DL logic. The abstract contract is a tupel 
\[(\vars,\; \plhs,\; \pre^a,\; \post^a_{norm},\; \mymod^a,\; \mathit{term},\; \defs)\] consisting of

\begin{align*}
  \vars &= \{\hPre,\; \self,\; \res,\; \exc\}\\
  \plhs &= \{\mathit{computeGradeR}, \mathit{computeGradeE}, \mathit{computeGradeA}, \mathit{env1}, \mathit{env2}\}\\
  \pre^a &= \begin{aligned}[t]
      &\mathit{computeGradeR}(\theap,\; \self) \\
      & \wedge \mathit{env1}(\theap,\; \self) \wedge \mathit{env2}(\theap,\; \self)
       \end{aligned}\\
  \post^a_{norm} &= \begin{aligned}[t]
      &\mathit{computeGradeE}(\theap, \hPre, \self, \res)\\
      &\wedge \mathit{env1}(\theap,\; \self) \wedge \mathit{env2}(\theap,\; \self)\\
      & !\mathit{exc} \keyeq \mathit{null}
       \end{aligned}\\
   \mymod^a &= \mathit{computeGradeA}(\hPre,\; \self)\\
   \mathit{term} &= \mathit{total}\\
   \defs &= \begin{aligned}[t]
   		&\{(\mathit{computeGradeR}, \mathit{def}_{\mathit{computeGradeR}}), (\mathit{computeGradeE}, \mathit{def}_{\mathit{computeGradeE}}),\\
   		&(\mathit{computeGradeA}, \mathit{def}_{\mathit{computeGradeA}}), 
		(\mathit{env1}, \mathit{def}_{\mathit{env1}}), (\mathit{env2}, \mathit{def}_{\mathit{env2}})\}
		 \end{aligned}
\end{align*}
where
\begin{align*}
	\mathit{def}_{\mathit{computeGradeR}} &= \self.\mathit{bonus} >= 0\\
	\mathit{def}_{\mathit{computeGradeE}} &= \res \keyeq \self.\mathit{exam} + \self.\mathit{bonus}\\
	\mathit{def}_{\mathit{computeGradeA}} &= \dot{\emptyset} \\
	\mathit{def}_{\mathit{env1}} &= \self.\mathit{exam} >= 0 \wedge \self.\mathit{bonus} >= 0 \wedge \self.\mathit{passingGrade} >= 0\\
	\mathit{def}_{\mathit{env2}} &= \self.\mathit{labs.length} \keyeq 10
\end{align*}

The terms $\mathit{def{\_}}$ result from the translation of JML expression appearing after the equality sign in the definition clauses into the \java{Java Card DL} logic. 

\subsection{Using Abstract Method Contracts}
The advantage of our definition of an abstract contract is that it can be used exactly as a concrete contract by an unmodified version of the \java{useMethodContract} rule. It guarantees the soundness of our abstract approach, provided the calculus using concrete contracts was sound.

\begin{center}
$\seqRule{}{\sequent{} {\{u\}\{w\}(\pre)}\\ 
\sequent{}{\{u\}\{w\}\{\hPre = \theap\}\{v\}(\post \rightarrow \dlbox{\java{r} = \res; \omega}\phi}}
{\sequent{}{\{u\}\dlbox{\java{r} = o.\java{m}(p_1, \dots p_n);\ \omega}\phi}}$
\end{center}

where $w = (\self := o\ ||\ \java{p}_1 := p_1\ ||\ \dots \ ||\ \java{p}_n := p_n)$ and $v = (\theap := anon(\theap, \mymod, h')\ ||\ \res : = a)$. 

The schema variables $\pre$ and $\post$ are substituted by the components $\pre^a$ and $\post^a_{norm}$ of an abstract method contract. The abstract modifies clause $\mymod^a$ appears in the anonymisation update in place of $\mymod$. The \java{Java Card DL} formulas with placeholders resulting from the premisses of the \java{useMethodContract} rule can be simplified by the calculus rules in the same way as normal ones. Among others, the symbolic execution of the programs in the modalities may be continued, as well as separation of the proof goals and even closure of some of them. However, it is impossible to close the goals that rely on the information that would otherwise be provided by the concrete clauses of a method contract. Furthermore, the anonymisation of the heap with an abstract modifies clause $\mymod^a$ can be compared to the full anonymisation of the heap until the placeholders in $\mymod^a$ are expanded with their definitions. It can be explained by showing the exact semantics of the $\mathit{select}$ operation on a small example from the proof using an abstract contract of the \java{computeGrade()} method. Let the term 
\[\mathit{select(anon(\heap,\ computeGradeA(\hPre,\ \self), h'),\ o,\ f)}\] 
be part of an open goal. Then it can be simplified as follows:
\[\mathit{if\ (o,\ f)} \in \mathit{computeGradeA(\hPre,\ \self)}\ \mathit{then\ (select(h',\ o,\ f))\ else\ (select(\heap,\ o,\ f))}\] 
The $\mathit{if\ (o,\ f)} \in \mathit{computeGradeA(\hPre,\ \self)}$ part cannot be resolved, because the set of assignable locations is represented by the placeholder. It is semantically equivalent to the situation, when the $\mathit{select}$ operation returns the value of \java{o.f} from the fresh heap $\mathit{h'}$.

The correctness proof for method \java{m}, which uses only abstract method contracts of method \java{m} itself and of all other methods that \java{m} invokes, is independent from the concrete specifications of these methods, that is, from the definitions given to placeholders. We refer to it as an \emph{abstract proof} or \emph{partial proof}. To prove that method \java{m} adheres to its concrete specification we have to expand the placeholders with their definitions. For this, every pair of the set $\defs$ is translated into a rewrite rule, which is added in the calculus as an axiom. For example, a rewrite rule corresponding to the pair $(\mathit{computeGradeE}, \mathit{def}_{\mathit{computeGradeE}})$ from the abstract contract of the \java{computeGrade()} method has the following form:
\begin{multline*}
\mathit{computeGrade(sv\_heap,\ sv\_heap}^\pre,\ \mathit{sv\_self,\ sv\_res)} \leadsto \\
\mathit{sv\_res \keyeq (select(sv\_heap,\ sv\_self,\ exam) + select(sv\_heap,\ sv\_self,\ bonus))}
\end{multline*}
where $\mathit{sv\_heap,\ sv\_heap}^\pre,\ \mathit{sv\_self,\ sv\_res}$ are schema variables and symbols $\mathit{exam}$ and $\mathit{bonus}$ are constant functions of type $\mathit{Field}$.

Before the expansion of the definitions takes place, we save the partial proof of method \java{m}. As long as the implementation of method \java{m} and the abstract clauses of the method contracts used in \java{m} have not changed, it can be loaded and reused to prove correctness of programs that have alternative concrete specification to the one we started with. More on the flexibility that abstract contracts offer is said in the next paragraph.

\paragraph{Flexibility of abstract method contracts}
Let there exist a partial proof for a method \java{m} that makes use of an abstract contract of the method \java{n}. Let's assume we wish to change the behavior of the method \java{n} and probably even of the class it belongs to. What are we allowed to alternate in the implementation and specifications of our program so that the existing partial proof could still be reused in the new version? The answer to this question is given by the following list. 
\begin{itemize}
	\item We can change the implementation of the method \java{n}.
	\item We can redefine placeholders both of the method \java{n} and the method \java{m}, thus completely changing the essence of their specifications.
\end{itemize}
However, there are some aspects on which the partial proof is solidly based and that cannot be changed without making it non-reusable:
\begin{itemize}
	\item The number, the types and the names of placeholders featured in all used method contracts
	\item Signatures of the methods \java{m} and \java{n}
	\item Type of the behavior of the methods \java{m} and \java{n} (normal or exceptional).
\end{itemize}

Nevertheless, even with these restrictions, the abstract method call approach offers a high level of flexibility that is exceptionally valuable when working on either several version of one program at a time or undertaking unexpected changes to already verified programs.

\chapter{Implementation}\label{ch:implementation}

As part of this thesis, we have extended the \KeY\ verification system to support abstract method calls. The main tasks were to implement the automatic translation of abstract method contracts from JML specifications into the \KeY\ logic and to provide the basic interface necessary for the workflow involving abstract method contracts, i.e construction, saving and reuse of a partial proof.
Our implementation is based on the \KeY\ system version 2.1. In this chapter we describe in details the changes to the source code that were required. We start in Section~\ref{sec:verifying-with-key} by giving a stepwise explanation of how a Java program is verified in \KeY\ in a general case and what parts of the \KeY\ architecture are responsible for each of the steps. 
In Section \ref{sec:foreword} we give a conceptual overview of our design of abstract method contracts and abstract invariants in the \KeY\ system. In Sections~\ref{sec:translation}~--~\ref{sec:save-and-load} we closely follow the steps described in Section~\ref{sec:verifying-with-key} and provide information on how we have extended or modified the implementation of the \KeY\ system to integrate the support of abstract method contracts and invariants.

\section{Verifying with \KeY}\label{sec:verifying-with-key}
In this section we consider a typical user story supported by \KeY\ system and describe which steps are taken to accomplish this task and how they are managed by the \KeY\ system. 

Let's assume that a user has implemented a Java program, has annotated it with JML and wishes to prove the correctness of program's methods with regard to the provided specifications. He loads the source files and as the first step, the Java source code is analyzed, which is handled by the package \java{java}. At this point the classes, their fields and methods are extracted and registered as such, the type hierarchy is extended.

At the next step, the JML specifications are extracted, which is handled by the package \java{speclang}. 
The main responsibility of the \java{JMLSpecExtractor} class is to extract either \emph{class level specifications} (mainly class invariants) for a given class or \emph{method level specifications} (mainly method contracts) for a given method. For this it first calls the preparser which is defined by the grammar \java{KeYJMLPreParser.g}. Its main task is to recognize different types of JML annotations and translate them into corresponding \emph{textual representations} (class \java{TextualJMLConstruct}). For example, a method contract returned by the preparser is represented by an object of the class \java{TextualJMLSpecCase} and its constituents (i.e. clauses, behavior modifies) are kept in this object individually as strings. 

After that, with help of the classes \java{JMLSpecFactory} and the \java{JMLTranslator}, the given textual constructs are transformed into \emph{specification elements} (class \java{SpecificationElement}) -- the internal representation of JML specifications in \KeY. The class \java{JMLTranslator} which makes use of the parser defined by the grammar \java{jmlparser.g} is responsible for translating textual JML expressions into correct terms (class \java{Term}) and the class \java{JMLSpecFactory} is responsible for the correct construction of specification elements. For example, a method contract is represented by the class \java{FunctionalOperationContract} and class invariants are represented by the class \java{ClassAxiom}. All specification elements are stored in an instance of the \java{SpecificationRepository} class.

In case of method contracts the class \java{JMLSpecExtractor} performs several preparatory steps on textual constructs before handing them over to the \java{JMLSpecFactory}. It means that based on the information already available at this time (characteristics of a program method, behavior type of a contract) it resolves the sugaring of JML and adds implicit specifications to the contract. Among others these are implicit non-null conditions for all method parameters, invariants of the class and default versions of omitted clauses. 

This concludes the translation and at the next step the user is presented with the list of available contracts to prove. 
When one of the contracts is chosen to be proven, the class \java{ProblemInitializer} builds a \emph{proof obligation} from it. Most of the preparatory work here is done by the class \java{AbstractOperationPO} (the prefix \emph{Abstract-} in the name refers to this class being abstract and has no connection to abstract contracts or abstract method calls) that prepares the program method, pre- and postcondition, declares necessary program variables and registers proof specific rules, for example relevant class invariants.

After the problem is initialized the user can attempt to prove it. The \KeY\ system allows for interactive or automated construction of proofs. Strategy for the automated proof search is defined in the class \java{JavaCardDLStrategy}, where every rule set is assigned with the cost for its application. The behavior of this basic search strategy can be modified with \emph{macros} that are designed for more specific search patterns. For example, the \java{Finish Symbolic Execution} macro puts priority on the symbolic execution of program in modalities and stops the search the moment the proof has no modalities left in the goals.

At any stage during the proof search or after it is possible to save the current state of the proof to the \java{.proof} file, which is managed by the class \java{ProofSaver}. The \java{.proof} file contains the name of the contract, the path to the java source file from which the given contract originated, the settings in which its proof was constructed and the tree of rule applications. To load the proof the \KeY\ system first translates the the provided java source file, then finds the right contract, constructs a corresponding proof obligation and repeats the listed rule applications on it. 

\section{Implementation vs Definition}\label{sec:foreword}
In this section we discuss the differences between our formal definitions of abstract invariants and method calls and their actual implementation.

First of all, even though we have provided the definition of abstract invariants in a form that completely resembles the approach used to define abstract contracts, the realization of the first in the \KeY\ system is completely different. Every class is assigned with the model field $\mathit{inv}$ representing its invariant. The JML invariants written by the user for one class are combined and translated into a \emph{represent clause} for the class' $\mathit{inv}$ model field. The model fields are used in contracts to denote the invariant in pre- and postcondition and when required, they are substituted by their representation through the application of a \java{useClassAxiom} rule. Postponing the application of this rule has the same effect as postponing the application of placeholder expansions. This design contributes to our motivation behind the abstract method call approach and because it was present in the implementation of the \KeY\ system before the start of our work, it has saved us the additional effort.

In contrast to our definition, the implementation of abstract method calls allows for the use of \emph{mixed} contracts. It means that not necessarily all clauses have to be abstract, instead the user himself chooses which clauses should be used abstractly by declaring appropriate placeholders and giving definitions to them. Such an approach is advantageous in cases, when parts of the specification are expected to stay the same in different program versions, because as we later show in Chapter~\ref{ch:evaluation}, full abstraction may lead to a significant increase of the proof size. Its drawback is, that an additional level of control is required from the user, because any omitted or implicitly implied by the JML sugaring features are interpreted in a concrete way. For example, if no requires clause or requires placeholder is provided, the contract will be completed with a concrete \requires\ \java{true} precondition. 

\section{Translation}\label{sec:translation}

The preparser recognizes the strings that declare placeholders (for example \rabs\ R) and stores them in an instance of the \java{TextualJMLSpecCase} together with requires, ensures or assignable clauses, depending on the keyword. The definition clauses strings are also stored in an instance of \java{TextualJMLSpecCase}, but in a separate container.

The parser deals with declaration of placeholders by creating a function or predicate according to Definition \ref{def:placeholder}, registering it in a namespace, instantiating its parameters with correct program variables and returning it as a term to the \java{SpecFactory}. These terms are used by \java{SpecFactory} to build clauses and subsequently to form an object of \java{FunctionalOperationContract} class. Translated method contracts are stored in an instance of the \java{SpecificationRepository} class.

For every definition clause the parser creates an instance of the \texttt{Abstract\-Contract\-Definition} class. It consists of a placeholder with instantiated parameters (field \java{definedSymbol}) and the term which defines it (field \java{definition}). It is checked during translation that types of the placeholder and its definition match. The \java{definition} term results from the translation of a JML expression. The instances of \java{AbstractContractDefinition} class are also stored in an instance of the \java{SpecificationRepository} class.

Overall, our changes were focused in the preparser and parser. In addition to that, we call to the translation of \java{AbstractContractDefinitions} inside the \java{extractMethodSpecs(\dots)} method of the \java{JMLSpecExtractor} class. But apart from that, no changes where done to the way \java{JMLSpecExtractor} or \java{JMLSpecFactory} handle the translation of method contracts.

\section{Rule Base}

The next step at which the modifications were required is the initialization of a proof obligation. As mentioned in Chapter~\ref{ch:amc}, according to our model the definitions of placeholders are integrated into \KeY\ as calculus rules or, to be more exact, as rewrite taclets. These are built from the instances of the class \java{AbstractContractDefinition} using the \java{TaclteGenerator} (see \java{AbstractContractDefinition.toTaclet(\dots)}). All taclets created from placeholder definitions belong to the rule set \java{"expand\_def"} and carry the names \java{"expand\_def\_}\textit{placeholderName}\java{"}. In the heuristic of the \java{JavaCardDLStrategy} they have the application cost of $-5700$, which puts them right after most basic rules of propositional logic (not, andLeft, orRight) and before any other simplification, splitting or java rules in the priority list. This is done to ensure that during a normal automated proof search they were applied as soon as possible. Placeholder definition taclets are added to the \java{JavaCardDL} calculus as axioms, which means that their application does not have to be justified.

The taclets for expansion of placeholders are loaded during the initialization of a proof obligation. More exactly, it happens in the \java{readProblem()} method of the \java{AbstractOperationPO} class, besides the loading of class axioms. 

\section{Construction of a Partial Proof}

By the time a user starts the construction of a proof for a chosen contract, the taclets for expansion of placeholders are already added to the rule base and available for interactive application. Moreover, calling the \KeY\ auto pilot will trigger their application and lead to a construction of a full concrete proof. To ensure, that an automated construction of a partial proof is possible, we have implemented a strategy macro under the name \emph{Finish Abstract Proof}. It serves as an additional filter between the standard strategy proposal of the next rule to be applied and its actual application on the goal. To serve our needs, this macro filters out the rules for the expansion of placeholders and for the usage of class invariants by setting their cost to infinity in the method \java{computeCost(\dots)} of the class \java{FinishAbstractProofMacro}. In any other sense the normal heuristic is used as defined by the \java{JavaCardDLStrategy} class. In particular, the work of the \java{Finish Abstract Proof} stops when no other rules can be applied, after which it is the right time to save the proof.

It has to be noted, that other macros or strategies were not modified, which means that during construction of the partial proof their usage should be avoided. Otherwise, they might use axioms and placeholder expansion rules which will break the idea behind the partial proof. Another aspect worth mentioning is that the standard \KeY's automated proof search strategy applied on a goal with placeholders (with expansion taclets turned off by the macro) tends to perform too many unnecessary proof splits trying to "guess" the value of placeholders or objects on the heap after anonymization. The best results in avoiding this can be achieved by turning off the \emph{proof splitting} in the \emph{JavaDL options} during the use of \emph{Finish Abstract Proof} macro.

\section{Saving and Loading a Partial Proof}\label{sec:save-and-load}
Caching and reusing the partial proof is the main point of our approach and necessary modifications to the \KeY\ system had to be made.

For the caching of the partial proof we have decided to use the existing functionality of the \KeY, namely its output format for the proofs. The form of the \java{.proof} file remains as described in Section~\ref{sec:verifying-with-key}. Saving of the proof can be done at any moment, but it makes sense to do it right after the \emph{Finish Abstract Proof} macro stops, as it indicates, that no further simplifications can be done. 

For the loading of the proof we reuse a lot of \KeY\ existing functionality, mainly presented in the \java{DefaultProblemLoader} class. Loading of the partial proof repeats the same steps as required for the normal loading of the proof, however, the path to the java source is taken not from the \java{.proof} file, but from the proof obligation for which the partial proof has to be loaded. This way, the newly constructed proof obligation is identical to the one that the user wants to prove, it contains the same taclets for the expansion of placeholders, but the rules applied to it are taken from the \java{.proof} file. Full replay of the cached proof on the new proof obligation is also advantageous as an additional check is made, whether the partial proof is compatible with the new proof obligation. 

The loading of the proof can be done by using the menu item \emph{Reuse proof} at the stage, when a proof obligation, for which we wish to reuse the partial proof, is opened and selected in the main window. After the replay of the partial proof the user can continue with construction of the concrete proof, using any of the macros and strategies available in the \KeY\ system.

\chapter{Evaluation}\label{ch:evaluation}

The objective of this chapter is to empirically evaluate the benefits in terms of the proof reuse offered by our implementation of abstract contracts in \KeY.
Our conjecture is that abstract contracts shall provide reasonable savings in the proof effort.

We conducted this empirical evaluation on two case study programs, where each of the programs come in several versions.
These program versions have a common top-level interface and differ in the implementation as well in the concrete specification of the called sub-routines.
The abstract specification of each sub-routine, however, stays the same.

This chapter is structured as follows.
In Section~\ref{sec:case-study-programs} we describe implementation and specification of our case study programs.
In Section~\ref{sec:experiments} we describe our experimental setup and elaborate on the nature of the undertaken experiments.
Section~\ref{sec:results} present the empirical results that we have obtained.
In Section~\ref{sec:interpretation} we interpret these results and derive a conclusion on the benefits offered by our implementation of abstract contracts.

\section{Case Study Programs}\label{sec:case-study-programs}

\subsection{{\tt Account} case study}
Our first case study is based on the example introduced in \cite{Thum12} and is a program consisting of two classes, \java{Account} and \java{Transaction}. It is implemented in five versions, of which we describe the first basic one (Listing~\ref{AccountV1}) in details and afterwards give the short description of differences between the versions.
\paragraph{Version 1} The class \java{Account} represents a simple bank account, with method \java{boolean} \java{update(int x)} for deposit or credit operation (depending on the sign of the parameter x) on the balance of the account. The reverse operation is offered by method \java{boolean} \java{undoUpdate(int x)}. Both methods signal whether the operation was successful with the boolean return value. In the first basic version these methods apply the change on the field \java{balance} unconditionally and always return \java{true}, which is also reflected in their contracts.\\

\begin{lstjavajml}[numbers = left, basicstyle=\small, caption=Implementation and specification of the classes \java{Account} and \java{Transfer} (Ver.~1), captionpos=b, frame = tb, label = AccountV1]
public class Account {
	int balance = 0;
	boolean lock = false;
	
	/*@ public normal_behavior
	 @ ensures (balance == \old(balance) + x) && \result; 
	 @ assignable balance;
	 @*/
	boolean update(int x) {
		balance = balance + x;
		return true;
	}
	/*@ public normal_behavior
	 @ ensures (balance == \old(balance) - x) && \result;
	 @ assignable balance;
	 @*/
	boolean undoUpdate(int x) {
		balance = balance - x;
		return true;
	}
	/*@ public normal_behavior
	 @ ensures \result == this.lock;
	 @*/
	boolean /*@ pure @*/ isLocked() {
		return lock;
	}	
}	

public class Transaction {
	/*@ public normal_behavior
	  @ requires destination != null && source != null;
	  @ requires source != destination;
	  @ ensures \result ==> (\old(destination.balance) + amount == destination.balance);
	  @ ensures \result ==> (\old(source.balance) - amount == source.balance);
	  @ assignable \everything;
	 @*/
	public boolean transfer(Account source, Account destination, int amount) {
		if (source.balance < 0) amount = -1;
		if (destination.isLocked()) amount = -1;
		if (source.isLocked()) amount = -1;
		
		int take;
		int give;
		if (amount != -1) { take = amount * -1; give = amount;} 
	
		if (amount <= 0) {
			return false;
		}
		if (!source.update(take)) {
			return false;
		}
		if (!destination.update(give)) {
			source.undoUpdate(take);
			return false;
		}
		return true;	
	}        
}    
\end{lstjavajml}

The account can be locked to prevent any operation over it, which is reflected by the field \java{boolean lock}. The state of the account lock is returned by  the method \java{boolean isLocked()}, which behavior is not carried among the versions. 

The methods of the class \java{Account} are used by method \java{boolean} \java{transfer} (\java{Account source}, \java{Account destination}, \java{int amount}) of the class \java{Transaction} to perform a transfer between two accounts. It takes the \java{amount} of money from \java{source} and pays it to \java{destination}. It performs a number of pre-checks to ensure that the transaction will be successful before actually performing the transfer. For instance, transfer operation is unsuccessful (and returns \java{false}) if either of the following conditions is met:
\begin{itemize}
	\item balance of the source account is negative
	\item one of the accounts is locked
	\item the transfer amount equals zero or is negative
	\item one of the \java{update} operations is negative.
\end{itemize}
Otherwise the method transfers money between the accounts and returns \java{true}. 

The contract of method \java{transfer(...)} in the basic version of the program specifies only the outcome of the successful transfer, more concretely, it should be proven that if the return value is \java{true}, then the new balances of accounts equal their old balances (before the call) plus (for \java{destination})/minus (for \java{source}) the value of the parameter \java{amount}.

\paragraph{Version 2} In the second version, the implementation of the classes remains unchanged. However, we simplify the contract of method \java{transfer} to such that it only has to be proven that if the transferable amount is negative, then the method returns \java{false}, which models one of the possible behaviors of this method. 

\paragraph{Version 3} In comparison with the first version of this program, the third version contains the notion of an overdraft limit (represented by the static field \java{int OVERDRAFT\_LIMIT} of the class \java{Account}). An additional test is implemented in the methods \java{update(\dots)} and \java{undoUpdate(...)}, to ensure, that the overdraft limit of the account is not exceeded as the result of the operation. When the operation is not possible, the method returns \java{false}.

On the side of the specification this modification of the code is reflected as follows. In the new ensures clauses of methods \java{update(...)} and \java{undoUpdate(...)} we perform a case distinction on whether the returned boolean result is true or false. In case the result is true, the specification coincides with the specification of the first version of our program. In case the result is false, the specification claims that the balance stays unchanged. The specification of the \java{transfer(...)} method is undisturbed by the overdraft limit test and repeats the one from Version 1.

\paragraph{Version 4}
The fourth version is a further extension of the third one. In addition to the overdraft limit, the account is restricted by the daily withdrawal limit, which sets the bound on the amount of money that can be withdrawn from the account in one day. The limit is represented by the static field \java{DAILY\_LIMIT} and the amount of money that was withdrawn in one day is accumulated in the field \java{int} \java{withdraw} of the class \java{Account}. The implementation of methods \java{update(...)} and \java{undoUpdate(...)} is extended to perform an additional limit test and to maintain the amount of money withdrawn in one day.

On the side of the specification this modification of the code required an adaptation of the ensure clauses of the methods \java{update(...)} and \java{undoUpdate(...)}. In the new ensures clauses of the methods \java{update(...)} and \java{undoUpdate(...)} we specify the state of the \java{withdraw} field after a successful or unsuccessful operation exactly the same way as it was done for the \java{balance} field in the third version. The specification of the \java{transfer(...)} method is undisturbed by the daily withdrawal limit test and repeats the one from  Version 1.
\paragraph{Version 5}
The fifth version is an alternative extension of the third one. In addition to the overdraft limit of the account, an update operation is now charged with a fee (represented by the static field \java{int FEE} of the class \java{Account}). It is subtracted from the \java{balance} during a successful \java{update(...)} operation and restored, when the method \java{undoUpdate(...)} is called.

On the side of the specification this modification of the code required an adaptation of the ensure clauses of the methods \java{update(...)} and \java{undoUpdate(...)}. In the parts of the ensures clauses that specify the state of the \java{balance} after a successful operation, the charge and restoration of the fee is taken into account. The specification of the \java{transfer(...)} method is adapted to accommodate the fee charge. Its postcondition is weakened from giving the exact values of the source and destination accounts after a successful transfer, to setting their lower bound.

The source code for all five versions of the \java{Account} case study can be found in the Section \ref{sec:a-account} of the Appendix.

\subsection{{\tt StudentRecord} case study}
The second case study for our evaluation is already introduced in Chapter 3 as the running example class StudentRecord. The method \java{passed()} is the top-level method of this program, which we verify in our experiments. This method calls the sub-routine \java{computeGrade()}. For the experiments we have additionally implemented a third simpler version of the class, presented in Listing~\ref{student3}\\
\begin{lstjavajml}[numbers = left, basicstyle=\small, caption=Implementation and specification of the class \java{StudentRecord} (Ver.~3), captionpos=b, frame = tb, label = student3]
class StudentRecord {
	// exam result
	int exam;
	
	// minimum grade necessary to pass the exam
	int passingGrade;
	
	// completed labs
	boolean[] labs = new boolean[10];
	
	//@ public invariant exam >= 0 && passingGrade >= 0;
	//@ public invariant labs.length == 10;

	/*@
        @ public normal_behavior
        @ requires true;
        @ ensures \result == exam;
        @ assignable \nothing;
        @*/
        int computeGrade(){
            return exam;
        }
	
	/*@
	@ public normal_behavior
	@ ensures \result ==> exam >= passingGrade;
	@ ensures \result ==> (\forall int x; 0 <= x && x < 10; labs[x]);
	@ assignable \nothing;
	@*/
	boolean passed() {
		boolean enoughPoints = computeGrade() >= passingGrade;
		boolean allLabsDone = true;
		for (int i = 0; i < 10; i++) {
	 	   allLabsDone = allLabsDone && labs[i];
		}
		return enoughPoints && allLabsDone;
	}
}
\end{lstjavajml}
As we see, in this version the class has no notion of the bonus at all, which influences the invariants of the class, the specification and implementation of the sub-routine, and the specification of the top-level method. 

Not surprisingly, the proofs of method \java{passed()} in the three different versions expose the same pattern.
All of them start with a comparably small portion to reflect the semantics of the first two assignments, which closely relies on the specification of the method \java{computeGrade()}. Afterwards, the major part of the proof deals with establishing the correctness of the loop invariant, which we provide as a JML annotation beforehand. In the end, the rest of the method is analyzed in the state after the loop and subsequently the correctness of the postcondition is established.

The source code for all three versions of the \java{StudentRecord} case study can be found in Section \ref{sec:a-student} of the Appendix.

\section{Setup and Experiments}\label{sec:experiments}
In order to assess the the benefits offered by our implementation of abstract contracts in \KeY\ we conduct two experiment series which we describe below.
Both series use two case study programs (including all their corresponding versions) which we presented in Section~\ref{sec:case-study-programs}.

We conducted all our experiments on a Mac laptop with Intel Core i5 2.4 GHz processor and 4 GB RAM running the Mac OS X 10.7.5  and Java 7. We used our modification of the \KeY\ system, which was based on the version 2.1 and implemented as defined in Chapter \ref{ch:implementation}.

\subsection{Experiment Series 1}

The goal of this experiment series is to empirically analyze how much of the proof efforts can be saved by using our implementation of abstract contracts.

Originally, for each of the described case study programs we wanted to conduct two experiments: the first with completely abstract specification, and the second with concrete specifications.
Completely abstract specifications follow the concept defined in Listing 3.4.
We state the abstract precondition by \rabs\ R, the abstract postcondition by \eabs\ E, and the abstract modifies clause by \aabs\ A.
Concrete specification directly state the desired pre- and postconditions as well the modifies clause.
We presented the essentials of the concrete specifications for our two case study programs in Section~\ref{sec:case-study-programs}.

However, after running several trial experiments and gaining some experience we envisioned the third way of giving specifications to which we refer as \emph{partially abstract specifications.}
They differ from the abstract specification in one aspect: the modifies clause is given concretely.
Stating the modifies clause concretely, like e.g., \assignable\ foo,  allows \KeY\ to keep the information about all variables except the variable foo after the application of the method contract and thus reduces the required verification effort.

Thus, for each of the described case study programs we conduct three experiments which differ in three kinds of specifications we use: completely abstract specification, partially abstract specification, and concrete specifications.

In each experiment with completely abstract specification and partially abstract specification as the first step we derive a partial proof.
For this, we load one of the versions of each program (we use the first version) and turn off the proof splitting option in the \KeY's settings.
We run the automated verification process in an abstract mode (using the \emph{Finish abstract proof} macro), and by this obtain the partial proof that we then save for the future reuse while verifying the remaining versions of each program.
By examining the \KeY\ statistics dialog, we note down the number of proof nodes as well as the number of branches in the partial proof.
Next, we consecutively load each version of the programs to be verified, load the saved partial proof, and run the normal automated verification process.
Similarly, we examine and note down the number of proof nodes as well as the number of branches in each of these proofs.

In each experiment with concrete specifications we directly run the normal verification process on each program version and note down the number of proof nodes as well as the number of branches in each of the obtained proofs.

\subsection{Experiment Series 2}

The goal of this experiment series is to empirically analyze how big are the parts of the proofs in our case studies that can be done by symbolic execution.
In this experiment series we use only concrete specifications.
For each of the case study program and for each of their version we run the following experiment.

After loading a program into \KeY\ we, firstly, run symbolic execution until possible and note down the number of proof steps.
Secondly, we close the resulting first-order goals with the auto pilot and note down the overall number of proof steps.

\section{Results of Experiments}\label{sec:results}

\subsection{Experiment Series 1}

The results of the experiment series 1 are given in Table~\ref{tab:results-11} and Table~\ref{tab:results-12} for case studies 1 and 2, respectively.
The rows of the tables, except for the last ones, correspond to different program versions.
Columns II and V present the number of proof steps (alternatively called nodes) in partial proofs, i.e., proof parts that have to be done once for all version of the program due to the usage of abstract contracts.
Columns III and VI present the number of proof steps required to finish the partial proof for every particular version of the program. 
Columns IV and VII present the number of proof steps in complete proofs, computed by summing up the number of steps in the partial proof with the number of steps for the rest of the proof.
Column VIII presents the number of proof steps in a full proof that uses concrete specifications.
The number of branches in the proofs is given in brackets.

The row ``Total'' gives the total number of proof steps needed to verify the complete programs of each case study, i.e., all versions of the programs.
For the experiments
using concrete specifications the total equals the sum of the number of
steps for the different versions, i.e., $\mathit{total}=\sum_{i=1}^{n}
\mathit{full}_i$ with $n$ being the number of program versions and $\mathit{full}_i$ being the number of steps in the full proof for the version $i$. 
In case of the experiments using partially or completely abstract specifications, we
compute the total of rest part of the proofs for different versions and add the size of the partial proof only once, as it can be reused between program versions. The formula for this computation has the following form: $\mathit{total}=(\sum_{i=1}^{n}
\mathit{rest}_i)+\mathit{partial}$, where $n$ is the number of program versions, $\mathit{rest}_i$ is the number of steps in the rest part of the proof for the version $i$ and $\mathit{partial}$ is the number of steps in the partial proof. 

\begin{table}[h]
\begin{center}
\begin{tabular}{|c||c|c|c||c|c|c||c|}
\hline
	& \multicolumn{3}{|c||}{Completely abstract}				& \multicolumn{3}{|c||}{Partially Abstract}                          & Concrete 	\\ \hline
I.	& II.						& III.			& IV.								& V.								& VI.			& VII.									& VIII.		\\ \hline
	& Partial				& Rest 	 	& Full 					& Partial              					& Rest 	& Full							& Full	\\ \hline\hline
v1	& \multirow{5}{*}{1390 (55)} 	& 676 (2)  	& 2066 (57)    	& \multirow{5}{*}{1408 (55)} 	&  457 (0) 	&1865 (55)   		& 1035 (63)	\\ \cline{1-1} \cline{3-4} \cline{6-8} 
v2&                   				&   492 (2)			& 1882 (57)    		&                   									& 315 (0)  	&1723 (55)    		& 972 (63)	\\ \cline{1-1} \cline{3-4} \cline{6-8} 
v3	&       						& 	1045 (4)			  	& 2435 (59)   		&                           	 		& 488 (0)  	& 1896 (55)   						& 1352 (68)	\\ \cline{1-1} \cline{3-4} \cline{6-8} 
v4&                   				& 	1329 (4)			 	& 2719 (59)     		&                            			&  567 (0) 	& 1975 (55)   						& 1461 (68)	\\ \cline{1-1} \cline{3-4} \cline{6-8} 
v5&                   				&   1349 (6)				& 2739 (61)    		&                            			&  619 (2) 	& 2027 (57)   						& 1601 (69)	\\ \hline

Total & \multicolumn{2}{|c|}{6281}    		 		&         		& \multicolumn{2}{|c|}{3854}     	&         					& 6421	\\ \hline
\end{tabular}
\caption{Results of experiment series 1 for case study 1 (account example)}
\label{tab:results-11}
\end{center}
\end{table}

\begin{table}[h]
\begin{center}
\begin{tabular}{|c||c|c|c||c|c|c||c|}
\hline
	& \multicolumn{3}{|c||}{Completely abstract}				& \multicolumn{3}{|c||}{Partially Abstract}                            		& Concrete	\\ \hline
	& Partial 				& Rest  		& Full 		& Partial               	& Rest 	 	& Full 	& Full	\\ \hline
I.	& II.						& III.			& IV.			& V.					& VI.			& VII.		& VIII.		\\ \hline\hline	
v1	& \multirow{3}{*}{514 (9)} 	&  677 (16) 	& 1191 (25)    		& \multirow{3}{*}{519 (9)} 		& 626 (16)  	& 1145 (25)    		& 919 (21)	\\ \cline{1-1} \cline{3-4} \cline{6-8} 
v2	&       							& 1292 (31)  	& 1806 (40)    		&                           	 		& 1263 (31)  	&1782 (40)   		& 1419 (36)	\\ \cline{1-1} \cline{3-4} \cline{6-8} 
v3 &                   				& 495 (15) 	& 1009 (24)   		&                            			& 418 (15)  	& 937 (24)     		& 904 (22)	\\ \hline 
Total & \multicolumn{2}{|c|}{2978}    		 		&         		& \multicolumn{2}{|c|}{2826}     	&         					& 3242		\\ \hline
\end{tabular}
\caption{Results of experiment series 1 for case study 2 (student example)}
\label{tab:results-12}
\end{center}
\end{table}

\FloatBarrier
\subsection{Experiment Series 2}

The results of the experiment series 2 are given in Table~\ref{tab:results-21} and Table~\ref{tab:results-22} for case studies 1 and 2, respectively.
The rows of the tables correspond to different program versions.
Columns II present the number of proof step in the symbolic execution parts of the proofs.
Columns III present the number of proof steps in complete proofs, where in brackets the number of branches of the proofs are given.
Columns IV give the percentage of the symbolic execution part of the proof in complete proofs.
They are obtained by dividing the numbers in Columns III by the numbers in Columns II.

\begin{table}[h]
\begin{center}
\begin{tabular}{|c|c|c|c|c|c|c|}
\hline
	& Symbolic execution	& Full proof 	& Ratio 		\\ \hline
I.	& II.					& III.			& IV.			\\ \hline\hline
v1	& 842 (53) 			& 1041 (63)  	& 81\%    		\\ \hline
v2	& 818 (53)     			& 981 (63)  	& 83\%    		\\ \hline
v3    	& 1037 (56)                  	& 1361 (68)  	& 76\%    		\\ \hline 
v4    	& 1147 (56)                  	& 1471 (68)  	& 78\%    		\\ \hline 
v5    	& 1141 (56)                  	& 1655 (68)  	& 69\%    		\\ \hline 
\end{tabular}
\caption{Results of experiment series 2 for case study 1 (account example)}
\label{tab:results-21}
\end{center}
\end{table}

\begin{table}[h]
\begin{center}
\begin{tabular}{|c|c|c|c|c|c|c|}
\hline
	& Symbolic execution	& Full proof 	& Ratio 		\\ \hline
I.	& II.					& III.			& IV.			\\ \hline\hline
v1	& 442 (11) 			& 955 (21)  	& 46\%    		\\ \hline
v2	& 494 (11)     			& 1451 (36)  	& 34\%    		\\ \hline
v3    	& 410 (11)                  	& 853 (21)  	& 48\%    		\\ \hline 
\end{tabular}
\caption{Results of experiment series 2 for case study 2 (student example)}
\label{tab:results-22}
\end{center}
\end{table}

\newcommand\observation[1]{
\begin{center}
\vspace{1em}
\fbox{\begin{minipage}{.9\textwidth}\noindent%
\emph{#1}
\end{minipage}}
\end{center}
}

\section{Interpretation of Experimental Results}\label{sec:interpretation}

We carefully investigated the results of the two experiment series that we described so far in this chapter.
This allowed us to derive six observations about the concept of abstract contracts which we present in this section.

\FloatBarrier
\subsection{Observation 1}

\observation{Deploying the concept of abstract contracts allows one to reuse up to 82\% of the proof.
Partially abstract specifications, on average, allow to reuse 63\% of the proof.
Completely abstract specifications, on average, allow to reuse 53\% of the proof.}

\begin{figure}[h]
        \centering
        \begin{subfigure}[b]{0.5\textwidth}
                \includegraphics[width=\textwidth]{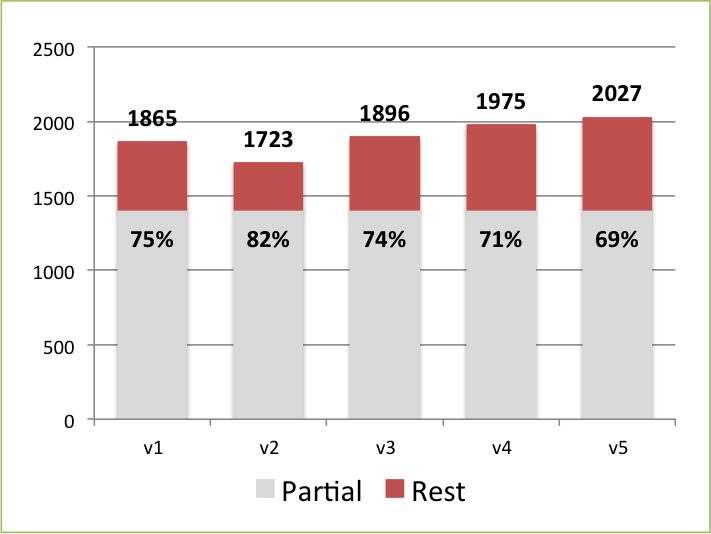}
                \caption{Partially abstract specification}
                \label{fig:account-partially-abstract}
        \end{subfigure}%
        ~ 
        \begin{subfigure}[b]{0.5\textwidth}
                \includegraphics[width=\textwidth]{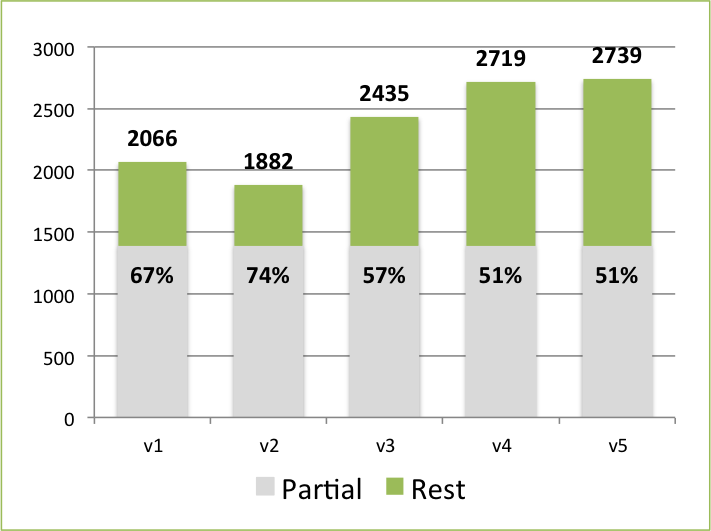}
                \caption{Completely abstract specification}
                \label{fig:account-completely-abstract}
        \end{subfigure}
        \caption{Proof reuse in case study 1 (account example)}
        \label{fig:obs1-account}
\end{figure}

\begin{figure}[h]
        \centering
        \begin{subfigure}[b]{0.5\textwidth}
                \includegraphics[width=\textwidth]{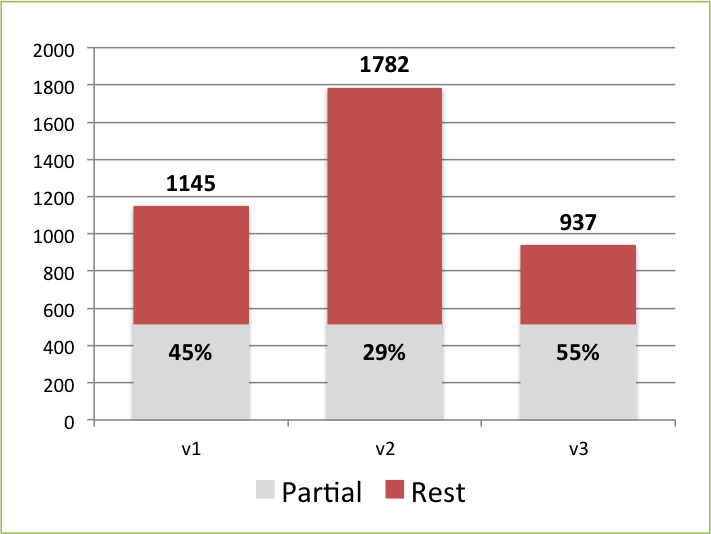}
                \caption{Partially abstract specification}
                \label{fig:student-partially-abstract}
        \end{subfigure}%
        ~ 
        \begin{subfigure}[b]{0.5\textwidth}
                \includegraphics[width=\textwidth]{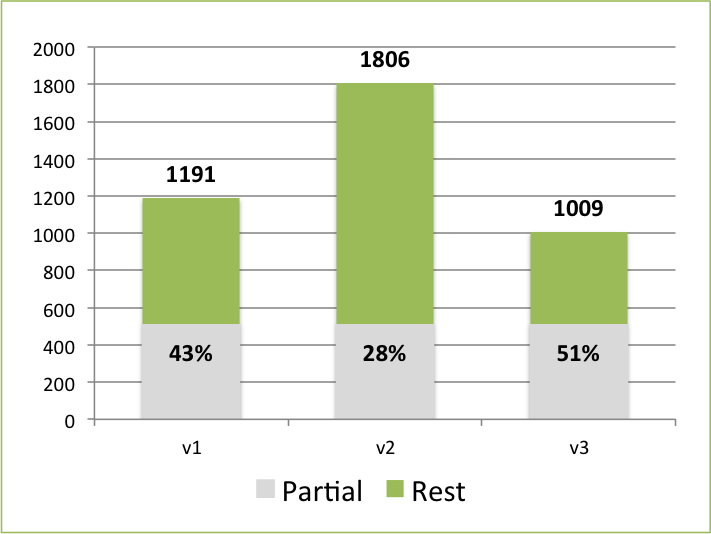}
                \caption{Completely abstract specification}
                \label{fig:student-completely-abstract}
        \end{subfigure}
        \caption{Proof reuse in case study 2 (student example)}
        \label{fig:obs1-student}
\end{figure}

This observation follows from analyzing Tables~\ref{tab:results-11} and~\ref{tab:results-12}.
Figures~\ref{fig:obs1-account} and~\ref{fig:obs1-student} depict the size of the proofs and percentages of proof reuse in our two case studies.
Figures on the left correspond to partially abstract specifications, whereas figures on the right correspond to completely abstract specifications.
The gray area corresponds to the partial proof that had to be constructed only once for all versions of the program.
The green area depicts additional steps that were required to finish the full proof for each version.
The percentage of reuse shows to what amount of the full proof the partial proof corresponds to, i.e how much of the full proof was done in single step by loading the pre-saved partial proof. 
The percentages are computed by dividing the size of the partial proof by the size of the full proof per program version.

It turns out that completely abstract specifications offer quantitatively worse proof reuse outcome than partially abstract specifications. However, they give more flexibility in alternation of the contracts content.
As discussed in Section~\ref{sec:experiments}, partially abstract specifications differ from completely abstract ones in that the modifies clause is given concretely in the former. 
This allows \KeY\ to keep the information about the heap at locations that are not listed in the modifies clause after the application of the method contract rule and as a consequence, close several branches in the partial proof itself. However, they restrict the user from changing the value of assignable clause in subsequent versions of the program.

\FloatBarrier

\clearpage
\subsection{Observation 2}
\vspace*{0.5cm}

\observation{The size of the proof reuse offered by abstract contracts is somewhat proportional to the size of the symbolic execution parts of proofs.}
\vspace*{0.5cm}

\begin{figure}[h]
        \centering
        \begin{subfigure}[b]{0.5\textwidth}
                \includegraphics[width=\textwidth]{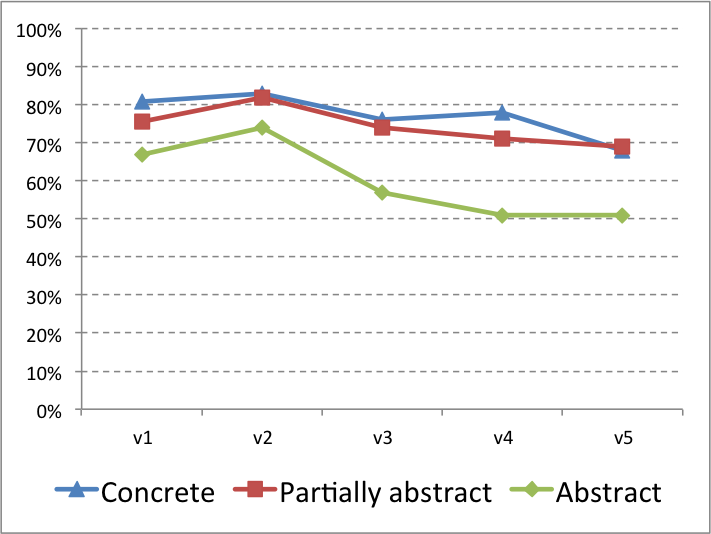}
                \caption{Account example}
                \label{fig:obs2-account}
        \end{subfigure}%
        ~ 
        \begin{subfigure}[b]{0.5\textwidth}
                \includegraphics[width=\textwidth]{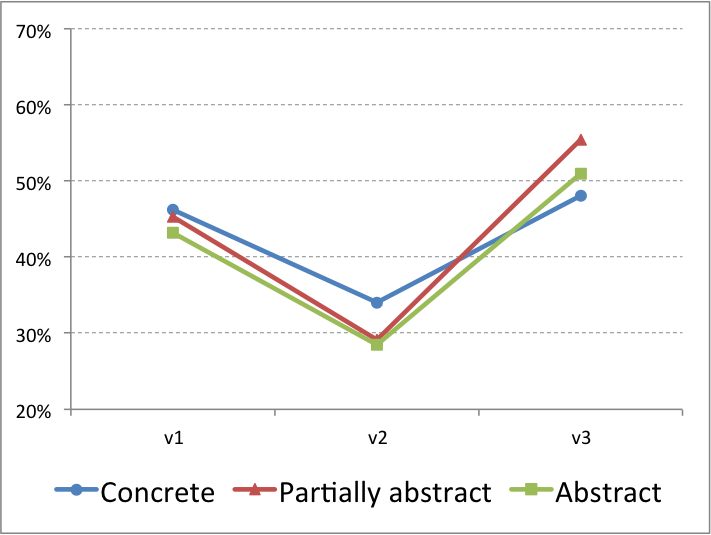}
                \caption{Student example}
                \label{fig:obs2-student}
        \end{subfigure}
        \caption{Percentage of reused proof (green and red) compared to the percentage of symbolic execution (blue)}\label{fig:obs2}
\end{figure}

This observation follows from comparing result tables of our first experiment series (Tables~\ref{tab:results-11} and~\ref{tab:results-12}) to the results tables of the second experiment series (Tables~\ref{tab:results-21} and~\ref{tab:results-22}).
We explored the relationship between the percentage of proof reuse offered by abstract contracts, both in their partially abstract and completely abstract versions, and the percentage of the proof steps that are done by means of the symbolic execution when verifying the same programs using concrete specifications.
We depicted these percentages for our two case studies in Figure~\ref{fig:obs2}.
In this figure, the red and the green curves corresponds to proof reuse which result from deploying partially and completely abstract specifications, respectively, whereas the blue curves correspond to the size of the symbolic execution part in proofs of the programs using concrete specifications.

It turns out that the behavior of the blue, red, and green curves is rather similar.
From this, one can conclude that the best proof reuse is achieved for programs with complex implementations but simple specifications.

\vspace*{1.5cm}

\FloatBarrier

\clearpage
\subsection{Observation 3}
\vspace*{0.5cm}

\observation{Deploying the concept of abstract contracts always increases the overall proof size per each program version. The size of such proofs involving completely abstract specifications is always larger than those involving partially abstract specifications.}
\vspace*{0.5cm}

\begin{figure}[h]
        \centering
        \begin{subfigure}[b]{0.5\textwidth}
                \includegraphics[width=\textwidth]{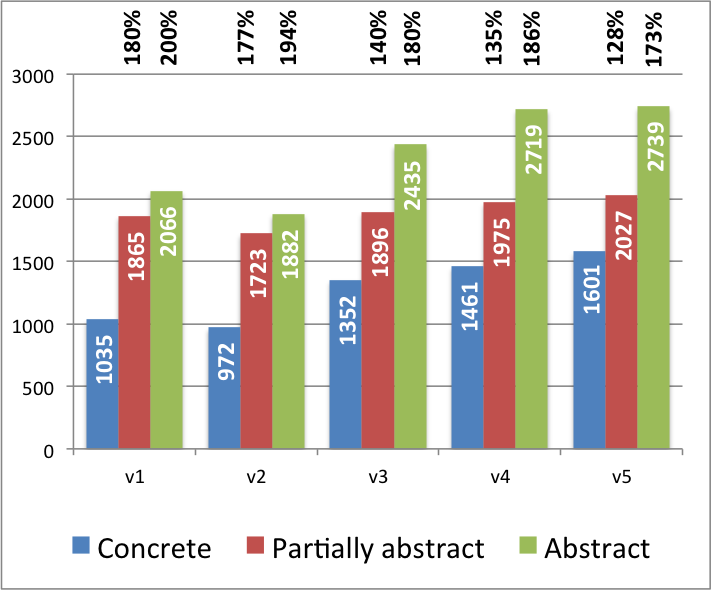}
                \caption{Account example}
        \end{subfigure}%
        ~ 
        \begin{subfigure}[b]{0.5\textwidth}
                \includegraphics[width=\textwidth]{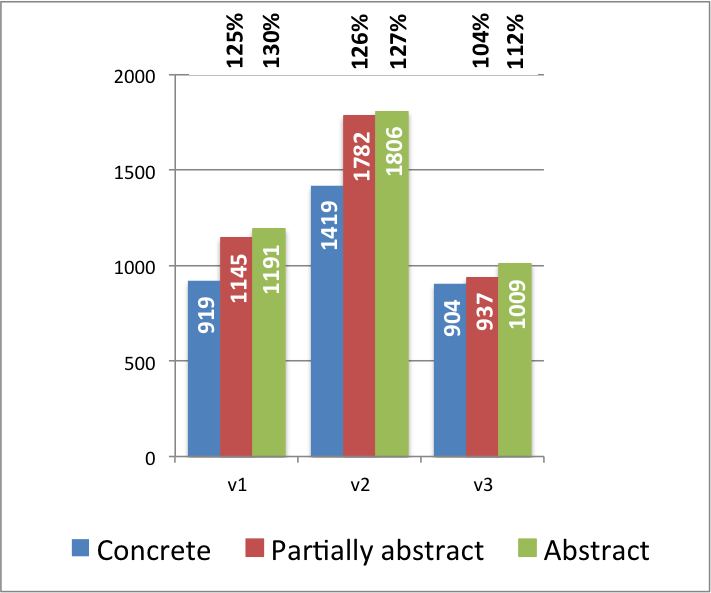}
                \caption{Student example}
        \end{subfigure}
        \caption{Proof size per version in nodes}
        \label{fig:obs3}
\end{figure}

This observation follows from considering the number of proof steps needed to verify each individual version of a program deploying concrete, partially abstract, and completely abstract specifications.
Figure~\ref{fig:obs3} depicts these numbers for our two case studies.
For each individual program version, we consider the number of required proof steps when verifying the program version against its concrete specification as 100\%.
Having this, using partially abstract specifications increases the proof effort for individual program versions by 39\% on average.
Using completely abstract specifications increases the proof effort for individual program versions by 62\% on average.

However as a partial proof can be reused between different program versions this additional proof effort will, as we will see in Observations 5 and 6, be compensated when considering the effort required for verification of all program versions together.
\vspace*{1.5cm}

\FloatBarrier
\clearpage
\subsection{Observation 4}
\vspace*{0.5cm}

\observation{Deploying the concept of abstract contracts almost does not increase the number of branches in a proof per each program version.}
\vspace*{0.5cm}

\begin{figure}[h]
        \centering
        \begin{subfigure}[b]{0.5\textwidth}
                \includegraphics[width=\textwidth]{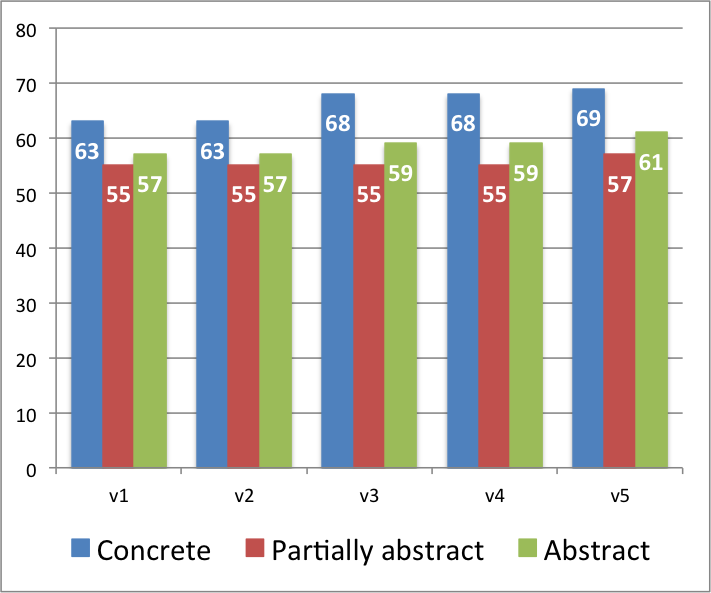}
                \caption{Account example}
        \end{subfigure}%
        ~ 
        \begin{subfigure}[b]{0.5\textwidth}
                \includegraphics[width=\textwidth]{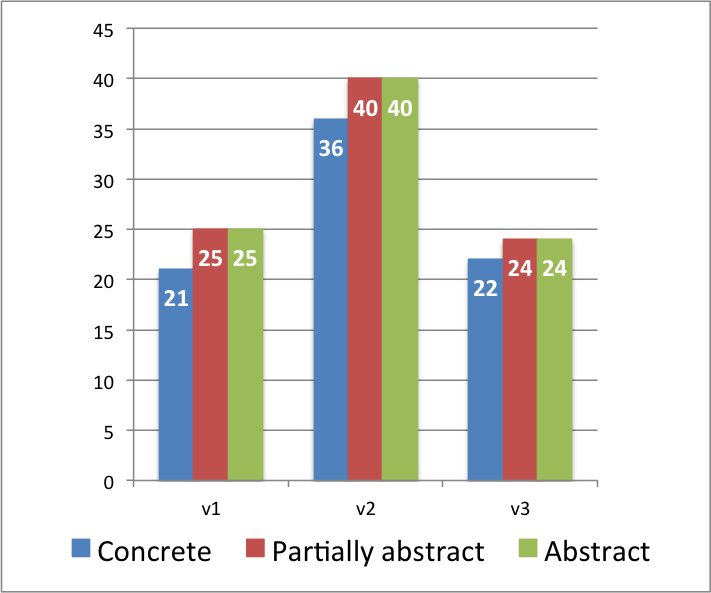}
                \caption{Student example}
        \end{subfigure}
        \caption{Number of branches in proof per version}
        \label{fig:obs4}
\end{figure}

This observation follows from considering the number of branches in full proofs for each individual version of a program built using concrete, partially abstract and completely abstract specifications. 
We depicted these numbers for both our case studies in Figure~\ref{fig:obs4}.
This observation means that the overall number of first-order goals that have to be closed to verify a program version is not affected by the deployment of abstract specifications. This is a great advantage, because the first-order goals can be delegated to SMT solvers, which usually offer a significant speed-up in closing such goals compared to the \KeY\ System. Assuming the development of interface between \KeY\ and SMT solvers, e.g. Z3, there is an expectation of rest parts of the proof taking insignificant effort in terms of time. 
\vspace*{2.0cm}

\FloatBarrier
\clearpage
\subsection{Observation 5}
\vspace*{0.5cm}

\observation{Deploying the concept of abstract contracts, both in partially and completely abstract flavors, does not enlarges the overall proof size per program, i.e., for all program versions together. The overall proof size per program is smaller for partially abstract contracts than for completely abstract contracts.}
\vspace*{0.5cm}

\begin{figure}[h]
	\centering
	\includegraphics[width=0.6\textwidth]{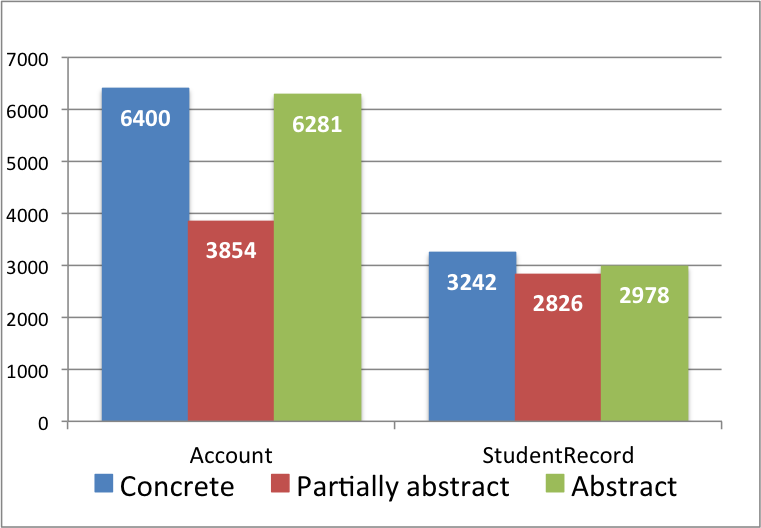}
	\caption{Total proof size in nodes for all versions per example}
	\label{fig:obs5}
\end{figure}

We compared the overall proof effort that is required for verification of each of our two case studies using concrete, partially abstract, and completely abstract specifications.
The result of this comparison is depicted in Figure~\ref{fig:obs5}.
Clearly, using partially abstract as well as fully abstract specifications does not increase the overall proof effort for verification of all versions of each of the programs together.
In fact, both partially and completely abstract specifications offered a decrease in the number of overall required proof steps, whereas for the abstract specifications this decrease is rather marginal and for the partially abstract specifications, as the account case study suggests, can be significant.

\vspace*{2.0cm}

\FloatBarrier
\clearpage
\subsection{Observation 6}

In order to derive the last observation we came up with an additional measurement notion which we call the \emph{amortized proof effort}. Because the partial proof is reused between versions of a program it make sense to distribute its size between all versions in order to assess the reuse benefit in evaluation of the proof effort per version. The amortized proof effort is measured in the number of proof steps per version. For proofs using completely abstract and partially abstract specifications it is computed by the formula $\mathit{APE}_i=\frac{\mathit{partial}}{n} + \mathit{rest}_i$, where $n$ is the number of program versions, $\mathit{partial}$ is the number of steps in the partial proof and $\mathit{rest}_i$ is the number of steps in the rest of the proof for the version $i$. For proofs using concrete specifications it equals the full size of the proof, i.e., $\mathit{APE}_i=\mathit{full}_i$, where $\mathit{full}_i$ is the number of steps in the full proof for the version $i$. We present these numbers in Table \ref{tab:results-31} and Table \ref{tab:results-32} for case studies 1 and 2 respectively.

\begin{table}[h]
\begin{center}
\begin{tabular}{|c||c|c|c|}
\hline
   & Completely abstract & Partially abstract & Concrete \\ \hline
v1 & 954                 & 738                & 1035     \\ \hline
v2 & 770                 & 596                & 972      \\ \hline
v3 & 1323                & 769                & 1352     \\ \hline
v4 & 1607                & 848                & 1461     \\ \hline
v5 & 1627                & 900                & 1601     \\ \hline
\end{tabular}
\caption{Amortized proof effort per version (account example)}
\label{tab:results-31}
\end{center}
\end{table}

\begin{table}[h]
\begin{center}
\begin{tabular}{|l|c|c|c|}
\hline
   & \multicolumn{1}{l|}{Completely abstract} & \multicolumn{1}{l|}{Partially abstract} & \multicolumn{1}{l|}{Concrete} \\ \hline
v1 & 848                                      & 799                                     & 919                           \\ \hline
v2 & 1463                                     & 1436                                    & 1419                          \\ \hline
v3 & 666                                      & 591                                     & 904                           \\ \hline
\end{tabular}
\caption{Amortized proof effort per version (student example)}
\label{tab:results-32}
\end{center}
\end{table}

Thus, it allowed us to derive the following observation.

\observation{The amortized proof effort of abstract contracts is always not larger than the one of concrete contracts. It is considerably smaller for partially abstract specifications and is comparable for fully abstract specifications.}

\begin{figure}[h]
        \centering
        \begin{subfigure}[b]{0.5\textwidth}
                \includegraphics[width=\textwidth]{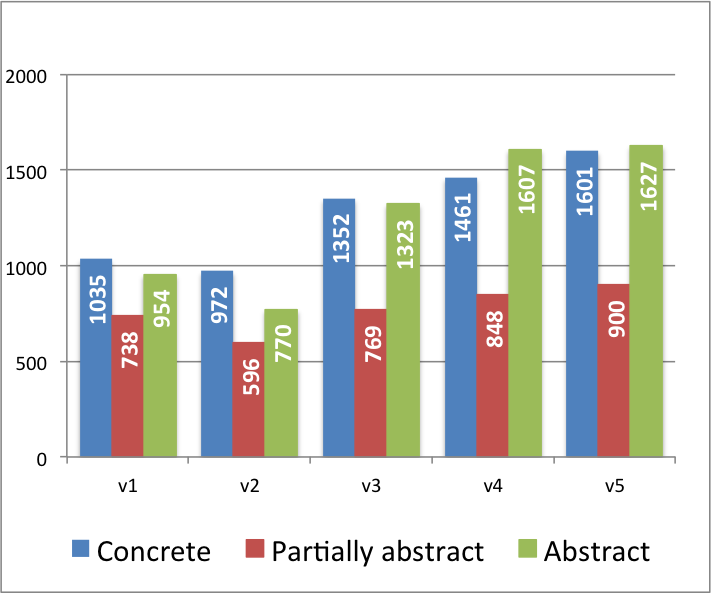}
                \caption{Account example}
        \end{subfigure}%
        ~ 
        \begin{subfigure}[b]{0.5\textwidth}
                \includegraphics[width=\textwidth]{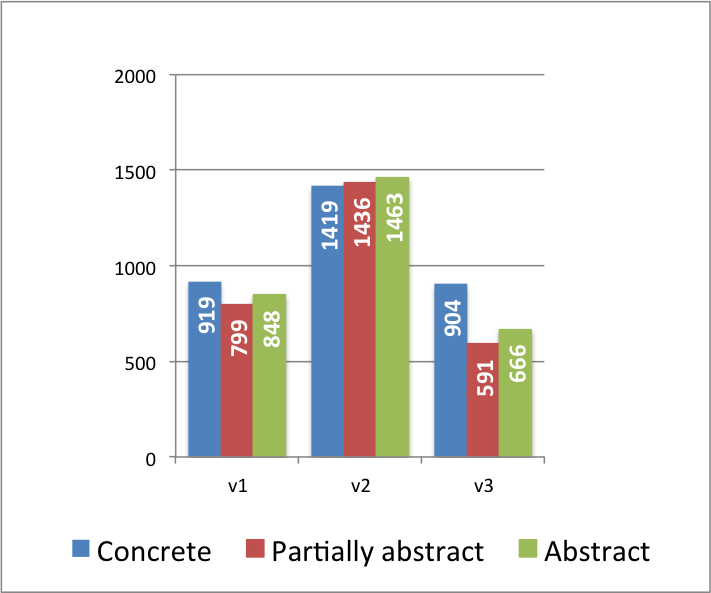}
                \caption{Student example}
        \end{subfigure}
        \caption{Amortized proof effort per version in nodes}
        \label{fig:obs6}
\end{figure}

This observation follows from comparing the size of the amortized proof effort for each individual version of a program deploying concrete, partially abstract, and completely abstract specifications.
Figure~\ref{fig:obs6} depicts these numbers for our two case studies.
Clearly, distributing the size of a partial proof among versions shows the gain of proof reuse, especially for the Account example.
This allows us to make an conjecture, that the benefit of deploying abstract contracts increases for the programs with higher number of versions.

\FloatBarrier

\section{Additional Experiments}\label{sec:additional-experiments}
The programs on which we have conducted our experiments so far have one drawback in the context of our work: the list of locations assignable by the subroutine methods was stable among different versions of the programs. On the one hand it has allowed us to investigate and compare the use of both completely and partially abstract contracts. On the other hand, it did not fully covered the achievements of this work, namely the introduction of fully abstract contract with abstract assignable clauses.
Moreover, the frame conditions that had to be proven for top-level methods were almost trivial. For example, the assignable clause for the method \java{transfer(\dots)} in the \java{Account} case study was specified as \everything. It meant that little proof effort was necessary after the expansion of assignable placeholders to close the goals.

To get the full picture of our work we decided to evaluate additional case studies that address these issues. In this section we proceed as follows. First of all we describe implementation and specification of our case study programs. Then we describe our experimental setup and present the empirical results that we have obtained. Finally, we interpret these results and derive a conclusion.

\subsection{Case Studies}
Our new case studies are based on the \java{Account} example introduced in Section~\ref{sec:case-study-programs}. The method \java{transfer(\dots)} of the class \java{Transaction} now not only performs the transfer of money between the accounts, but also writes the information about this transaction into the log.

The class \java{Log} is implemented in three versions. All of them model the log as an array \java{int[] logRecord} of integers and have a public method \java{void} \java{add}\java{(int bal)} and a private method \java{void} \java{rotateLog()}. The first one rotates the log to get a new empty slot and writes information to it, while the second one implements the rotation. There version of \java{Log} class differ in a way the log is rotated to get a new empty slot.

\paragraph{Ring}
In this version the log is modeled as a ring. New data is written into the next empty location of an array and if the array is full, the process starts from the beginning, overwriting the existing data. Hence, the assignable clause of method \java{add} has the following form:
\begin{center}
	\texttt{//@ \assignable\ logRecord[(last + 1) \% logRecord.length], last; }
\end{center} 
where \java{last} is a pointer to the last written location of the array.

\paragraph{Empty}
In this version new data is written into the next empty location of an array and if the array is full, all of the data it contains is deleted and the process starts at the beginning. Hence, the assignable clause of method \java{add} has the following form:
\begin{center}
	\texttt{//@ \assignable\ logRecord[*], last; }
\end{center} 
where \java{last} is a pointer to the last written location of the array.

\paragraph{Replace}
In this version new data is written into the next empty location of an array and if the array is full, it is discarded completely and the new empty array is created, in which the process starts at the beginning. Hence, the assignable clause of method \java{add} has the following form:
\begin{center}
	\texttt{//@ \assignable\ logRecord, last, logRecord[last + 1];}
\end{center} 
where \java{last} is a pointer to the last written location of the array.

The code for all three version of the \java{Log} class can be found in Section~\ref{sec:a-log} of the Appendix.

For the integration of class \java{Log} into the \java{Account} example we have chosen its most basic first version. An instance of class \java{Log} is included as a field into the \java{Transaction} class. Method \java{add(\dots)} is called in the \java{transfer(\dots)} method. Because the invocation of method \java{add(\dots)} has a strong effect on the current heap we have decided to investigate two possible situations. Once it is called at the very beginning of the method \java{transfer(\dots)}. It models that the information about the transaction is always written into the log independently of the fact, whether the transfer of money actually occurs. For the proof search it means, that all branches caused by the symbolic execution have to operate on the heap anonymised by the application of the contract of the method \java{add(\dots)}. We refer to this program as \emph{Begin} case study. In the second case study the method \java{add(\dots)} is called in one branch only, namely only after the transfer of money actually takes place. It means, that the major part of the proof is unaffected by the anonymisation of the heap by the application of the \java{add(\dots)} contract. We refer to this program as \emph{End} case study.

The frame condition for the method \java{transfer(\dots)} is much stricter, than it was in the previous experiments. Its assignable clause explicitly lists all modifiable locations, which directly copies the locations listed in the contract of a corresponding \java{add(\dots)} plus the \java{balance} fields of accounts taken as parameters.

\subsection{Setup and Results}
For each of the case studies we have provided completely abstract specifications and concrete specifications. The setup for the experiments using these two types of specifications repeats the one described in Section~\ref{sec:experiments}. For the proofs constructed using completely abstract specifications we note down the size of the partial proof and the full proof. For the proofs constructed using concrete specifications we note done the number of steps performed during symbolic execution and the size of the full proof. The size of the proof is expressed in terms of nodes and branches (given in parenthesis). 

The results for the \emph{Begin} and \emph{End} case studies are presented in the Table~\ref{tab:begin} and the Table~\ref{tab:end} respectively. The rows of the tables correspond to different versions of the \java{Log} class. Columns V. show the ratio of the size of the partial proof to the size of the full proof when using abstract specifications, that is the percentage of the reusable proof part in the full proof. Columns VIII. give the ratio of symbolic execution part of the proof to the full proof when using concrete specifications. Row ``Total'' shows the aggregated proof effort for all versions. It is computed using the formula defined in Section~\ref{sec:experiments}.

\begin{table}[h]
\begin{center}
\begin{tabular}{|c||c|c|c||c|c|c|}
\hline
			& \multicolumn{3}{|c||}{Completely abstract}						& \multicolumn{3}{|c|}{Concrete}	\\ \hline
			& Partial 							& Full 					& Ratio			& Symbolic execution 	& Full				& Ratio 		\\ \hline
I.			& II.									& IV.						& V.				& VI.								& VII. 				& VIII.		\\ \hline\hline	
Ring		& \multirow{3}{*}{1403 (58)} 	& 24012 (448)  & 6\%			& 1449 (56)					& 25303 (278)	& 6\%			\\ \cline{1-1} \cline{3-7} 
Empty	&       								& 10974 (301)    	& 13\%			& 1380 (56)					& 5777 (123)	& 24\%			\\ \cline{1-1} \cline{3-7}
Replace &                   					& 17060 (517)   	& 8\%			& 1346 (56)					& 16666 (485)	& 8\%		\\ \hline 
Total & \multicolumn{3}{|c||}{49240}          										&  \multicolumn{3}{|c|}{47746}		\\ \hline
\end{tabular}
\caption{Results for \java{Account+Log}, \emph{Begin} case study}
\label{tab:begin}
\end{center}
\end{table}

\begin{table}[h]
\begin{center}
\begin{tabular}{|c||c|c|c||c|c|c|}
\hline
			& \multicolumn{3}{|c||}{Completely abstract}						& \multicolumn{3}{|c|}{Concrete}	\\ \hline
			& Partial 							& Full					& Ratio 		& Symbolic execution 	& Full			& Ratio		\\ \hline
I.			& II.									& IV.						& V				& VI.								& VII.				& VIII.		\\ \hline\hline	
Ring		& \multirow{3}{*}{1583 (58)} 	& 4489 (111) 	& 35\%    		& 1206 (56)					& 2692 (86)	& 45\%		\\ \cline{1-1} \cline{3-7} 
Empty	&       								& 4003 (106)  		& 40\% 			& 1115 (56)     				& 2141 (79)	& 52\%		\\ \cline{1-1} \cline{3-7}
Replace &                   					& 4454 (114)  		& 36\%   		& 1118 (56)                  	& 2666 (91)	& 42\%		\\ \hline 
Total & \multicolumn{3}{|c||}{9780}          										&  \multicolumn{3}{|c|}{7499}		\\ \hline
\end{tabular}
\caption{Results for \java{Account+Log}, \emph{End} case study}
\label{tab:end}
\end{center}
\end{table}

\subsection{Interpretation}
The results of additional experiments give mixed impression. First of all, we notice that the reuse percentage has dropped dramatically to the average of 10\% in the \emph{Begin} example, while also falling under 40\% in the \emph{End} example. Secondly, we see that the aggregated proof effort required to verify all version of the program is no longer smaller when using abstract approach than when using concrete specification, which might be connected to the reuse percentage. Another interesting feature is that while the deployment of abstract contracts increases the proof effort pro program version compared to the concrete approach on average by 75\% in the \emph{End} case study, the \emph{Begin} case study shows reverse pattern. Two program versions, \emph{Ring} and \emph{Replace}, experienced almost no increase at all.

However, one of our previous observations got confirmed by the additional experiments. There exists a very strong correlation between the proof reuse percentage when using abstract specifications and the ratio between symbolic execution and complete proof when using concrete specifications. This leads us to an assumption, that abstract contracts are less effective in terms of proof saving efforts in context of programs that have complex specification and lead to first-order goals that are difficult to prove. In proofs for such programs the symbolic execution part, which we aim to reuse, becomes insignificant comparing to the first-order reasoning part. The situation is further enhanced when the program method contains many conditional statements or method calls that split the proof. As a result, a lot of identical simplification steps have to be repeated in different branches, especially if the placeholders are replaced with their definitions only after all splits were done. This particular issue could be tacked by recognizing of similar goals for which the same proof steps can be applied. The use of placeholders can help here because they hide the complex details of first-order goals and similar patterns get more recognizable.

\chapter{Related Work}\label{ch:relwork}

Since the release of the Java programming language in 1995, formal reasoning about correctness and security of Java programs attracted much attention by the scientific community and a lot of progress has been made.
Below we list some of the important milestones in verification of Java programs achieved over the last two decades.
First experiments in formalization and deductive verification of Java programs in the verification system PVS were done in the LOOP project~\cite{Jacobs98}.
Huisman and Jacobs~\cite{Huisman00} defined a set of Hoare logic rules for verifying Java programs in PVS, including the reasoning about abrupt termination.
While these were experiments in mechanical interactive verification, Havelund and Pressburger~\cite{Havelund00} proposed to apply the SPIN model checker for automatic verification of Java programs which were translated to the Promela language of SPIN with the Java PathFinder tool.
Subsequent developments in the area went in the directions of making verification tools for Java more practical and applicable in the industrial context.
This includes the start of the development of the \KeY\ tool~\cite{Ahrendt00} which is now ongoing for almost fifteen years and also has given the context for this thesis.
Another tool with the similar goal is, for instance, the jStar tool~\cite{Distefano08}.
Beside tools that aim at formal functional verification of Java programs, there is a line of tools that specifically address static verification of the security of Java programs (e.g.,~\cite{Myers99, Giffhorn08, Lux11}).

Regarding the reuse of proofs in software verification, this topic has attracted attention already in the beginning of the 1990's~\cite{Reif93} and a number of different approaches has been explored since then.
Felty and Howe~\cite{Felty94} prototyped a theorem proving system that supports generalization of tactic proofs by using meta variables, and by that offers possibilities for proof reuse.
Melis and Whittle~\cite{Melis99} investigate reasoning by analogy in the context of interactive theorem proving and suggested an approach how to reuse gathered problem solving experience in proof planning. 
Schairer and Hutter~\cite{Schairer02} explore how proofs evolve together with modifications of formal specifications of the software.
They introduce a set of basic transformations for specifications which induce the corresponding transformations of the proofs.
An incremental proof reuse mechanism based on similarity measure is proposed by Beckert and Klebanov~\cite{Beckert04} in the context of the \KeY\ project. 
This mechanism allows to reuse proof steps even if the situation in the new proof is not identical to the existing template.
Hutter and Autexier~\cite{Hutter05} discuss how to maintain the dependence between formal specifications and proofs in large formal software development projects by means of development graphs.
For each modification, the effect in the development graph is computed such that only invalidated proofs
have to be re-done. 
Bourke et al.~\cite{Bourke12} report on the challenges an the experience with proof reuse in two large-scale formal verification projects~\cite{Alkassar08,Klein09}.
Dovland, Johnsen, and Yu~\cite{Dovland12} introduce a set of allowed changes for evolution of object-oriented programs.
In their approach, a proof context is constructed which keeps track of proof obligations for verified method contracts.
Program changes cause the proof context to be adapted so that the proof obligations that are still valid are
preserved and new proof goals are created.

Recently H{\"a}hnle, Schaefer, and Bubel~\cite{Haehnle13} proposed the concept of abstract method contracts in order to facilitate proof reuse in software verification.
This work is directly extended by the current thesis.

\chapter{Conclusion and Future Work}\label{ch:conclusion}

In this thesis, we presented the implementation and evaluation of the concept of abstract operational contracts~\cite{Haehnle13} within the \KeY\ verification system.

The implementation part of this thesis delivered three technical contributions. Firstly, we extended the set of specification forms that may be defined abstractly with abstract locations sets and abstract class invariants. We provided supporting formal definitions and extended the JML to cover abstract specifications.
Secondly, we adapted the JavaDL logic for proper handling of abstract method calls.
Thirdly, we adapted verification workflow of the \KeY\ system to facilitate automatic reuse of proofs by clearly separating and caching reusable proof parts. We also implemented necessary changes to the graphical user interface and to the proof search strategies in the \KeY\ system.

In the evaluation part of this thesis we conducted several experiments in order to understand the benefits of the implementation of abstract contracts in a practical setup. This has revealed three main practical advantages of our implementation of abstract contracts.

\begin{description}
	\item{\bf Better modularity.} Abstract contracts offer a possibility to conduct a preliminary proof for a program even when no specification for this program is given, e.g., immediately right after the program has been written. This preliminary proof the can be reused during any proper verification of the program against any desired concrete specification.

	\item{\bf Better flexibility.} Our implementation of abstract contracts allows the user to modify more aspects of specifications without breaking the partial proof. In particular, our implementation allows the user to modify not only pre- and postconditions, but also program invariants as well as assignable clauses.

	\item{\bf Non-increase in number of proof branches.} Although using abstract contracts increases the total number of proof steps for the verification of a particular program version, it does not increase the number of proof branches. To this end, it looks promising to apply SMT solvers in order to discharge open first-order goals of the partial proofs.
\end{description}

Moreover, during our practical evaluation of abstract contracts, we gathered a number of interesting insights on how the shape of specifications and of programs under verification affect the resulting number of proof steps. In the table below, we summarize these observation by classifying whether a particular aspect has more positive or more negative effect on the number of proof steps.

\begin{table}[h]
\centering
\begin{tabular}{|l|l|}
\hline
\multicolumn{1}{|c|}{More positive}      & \multicolumn{1}{c|}{More negative}       \\ \hline \hline
concrete assignable clause               & abstract assignable clause               \\ \hline
complex program \& simple specifications & simple program \& complex specifications \\ \hline
long single execution path               & rich branching                           \\ \hline
\end{tabular}
\end{table}

\bigskip

Regarding the future work, among others, the two following directions offer themselves as meaningful candidates for an immediate follow-up to this thesis.

\begin{description}

	\item{\bf Increase of proof reuse.} Due to the abstract contracts there is a certain number of the generated identical subgoals  and, respectively, proof parts that can be handled in a more effective fashion. These identical subgoals appear when using the abstract contracts because, in contrast to using the concrete contracts, the \KeY\ system first performs the split step and then the simplification steps. As a consequence there is a numebr of identical simplification steps generated. Search for and recognition of such proof parts is a promising direction for further increase of proof reuse. 

	\item{\bf Integration of the approach.} Our implementation of abstract contracts can be integrated in the background verification plugin for Eclipse in order to support software engineers with an efficient, on-the-fly verification. Here, partial proofs can be constructed and saved automatically while the software engineer develops a program. Furthermore, any brake of the proofs due to modifications done in the subroutines can be detected, and the respective partial proof can be loaded automatically before any concrete proof steps are performed.
\end{description}

\bibliographystyle{alpha}
\bibliography{Pelevina14}

\appendix

\chapter{Appendix}\label{appendix}
\section{Code of the {\tt StudentRecord} example}\label{sec:a-student}
In this section of the appendix we provide the code of the \java{StudentRecord} example as it was used in our experiments from the Chapter Evaluation. For every version of the program we provide three Listings.
\begin{itemize}
	\item First listing corresponds to the experiments using concrete specification
	\item Second listing corresponds to the experiments using partially abstract specification
	\item Third listing corresponds to the experiments using completely abstract specification
\end{itemize}
Please note, that in the listings of the second and the third category, we only show the specification of the methods and omit their implementation, as the latter does not change.

\subsection{Version 1}\label{subsec:a-s-1}
\begin{lstjavajml}[captionpos=b, frame = single, caption={Concrete specification, Ver. 1, \java{StudentRecord}}, basicstyle=\footnotesize]
class StudentRecord {
	// exam result
	int exam;
	
	// achieved bonus
	int bonus;
	
	// minimum grade necessary to pass the exam
	int passingGrade;
	
	// completed labs
	boolean[] labs = new boolean[10];
	
	//@ public invariant exam >= 0 && bonus >= 0 && passingGrade >= 0;
	//@ public invariant labs.length == 10;

	/*@
    @ public normal_behavior
    @ requires bonus >= 0;
    @ ensures \result == exam + bonus;
    @ assignable \nothing;
    @*/
    int computeGrade() {
        return exam + bonus;
    }
	
	/*@
	@ public normal_behavior
	@ ensures \result ==> exam + bonus >= passingGrade;
	@ ensures \result ==> (\forall int x; 0 <= x && x < 10; labs[x]);
	@ assignable \nothing;
	@*/
	boolean passed() {
		boolean enoughPoints = computeGrade() >= passingGrade;
		boolean allLabsDone = true;
		/*@ loop_invariant 0 <= i && i <= 10 
	    		 && (\forall int x; 0 <= x && x < i; 
			 			allLabsDone ==> labs[x]);
	   	@ assignable allLabsDone;
	   	@ decreases 10 - i;
	  	@*/
		for (int i = 0; i < 10; i++) {
	 	   allLabsDone = allLabsDone && labs[i];
		}
		return enoughPoints && allLabsDone;
	}
}

\end{lstjavajml}
\bigskip
\begin{lstjavajml}[captionpos=b, frame = single, caption={Partially abstract specification, Ver. 1, \java{StudentRecord}}, basicstyle=\footnotesize]
class StudentRecord {
	{ ... } // Fields as in the Listing A.1
	
	//@ public invariant exam >= 0 && bonus >= 0 && passingGrade >= 0;
	//@ public invariant labs.length == 10;
    
    /*@
	@ public normal_behavior
	@ requires_abs computeGradeR; 
	@ ensures_abs computeGradeE;
	@ assignable \nothing;
	@ def computeGradeR = bonus >= 0;
	@ def computeGradeE = \result == exam + bonus;
	@*/
    int computeGrade() { ... }
	
	/*@
	@ public normal_behavior
	@ requires_abs passedR;
	@ ensures_abs passedE1;
	@ ensures_abs passedE2;
	@ assignable \nothing;
	@ def passedR = true;
	@ def passedE1 = \result ==> exam + bonus >= passingGrade;
	@ def passedE2 = \result ==> (\forall int x; 0 <= x && x < 10; labs[x]); 
	@*/
	boolean passed() { ... }
}

\end{lstjavajml}
\bigskip

\begin{lstjavajml}[captionpos=b, frame = single, caption={Completely abstract specification, Ver. 1, \java{StudentRecord}}, basicstyle=\footnotesize]
class StudentRecord {
	{ ... } // Fields as in the Listing A.1
	
	//@ public invariant exam >= 0 && bonus >= 0 && passingGrade >= 0;
	//@ public invariant labs.length == 10;

	/*@
	@ public normal_behavior
	@ requires_abs computeGradeR; 
	@ ensures_abs computeGradeE;
	@ assignable_abs computeGradeA;
	@ def computeGradeR = bonus >= 0;
	@ def computeGradeE = \result == exam + bonus;
	@ def computeGradeA = \nothing;
	@*/
    int computeGrade() { ... }
	
	/*@
	@ public normal_behaviour
	@ requires_abs passedR;
	@ ensures_abs passedE1;
	@ ensures_abs passedE2;
	@ assignable_abs passedA;
	@ def passedR = true;
	@ def passedE1 = \result ==> exam + bonus >= passingGrade;
	@ def passedE2 = \result ==> (\forall int x; 0 <= x && x < 10; labs[x]); 
	@ def passedA = \nothing;
	@*/
	boolean passed() { ... }
}

\end{lstjavajml}
\bigskip

\subsection{Version 2}\label{subsec:a-s-2}

\begin{lstjavajml}[captionpos=b, frame = single, caption={Concrete specification, Ver. 2, \java{StudentRecord}}, basicstyle=\footnotesize]
class StudentRecord {

	// exam result
	int exam;
	
	// achieved bonus
	int bonus;
	
	// minimum grade necessary to pass the exam
	int passingGrade;
	
	// completed labs
	boolean[] labs = new boolean[10];
	
	//@ public invariant exam >= 0 && bonus >= 0 && passingGrade >= 0;
	//@ public invariant labs.length == 10;
	
    /*@
    @ public normal_behavior
    @ requires bonus >= 0;
    @ ensures (exam >= passingGrade) ==> \result == exam + bonus;
    @ ensures (exam < passingGrade) ==> \result == exam;
    @ assignable \nothing;
    @*/

    int computeGrade(){
        if (exam >= passingGrade) {
            return exam + bonus;
        } else {
            return exam; 
        }}
	
	/*@
	@ public normal_behavior
	@ ensures \result ==> exam + bonus >= passingGrade;
	@ ensures \result ==> (\forall int x; 0 <= x && x < 10; labs[x]);
	@ assignable \nothing;
	@*/
	boolean passed() {
		boolean enoughPoints = computeGrade() >= passingGrade;
		boolean allLabsDone = true;
		/*@ loop_invariant 0 <= i && i <= 10 
	    		 && (\forall int x; 0 <= x && x < i; 
			 			allLabsDone ==> labs[x]);
	   	@ assignable allLabsDone;
	   	@ decreases 10 - i;
	  	@*/
		for (int i = 0; i < 10; i++) {
	 	   allLabsDone = allLabsDone && labs[i];
		}
		return enoughPoints && allLabsDone;
	}
}
\end{lstjavajml}
\bigskip

\begin{lstjavajml}[captionpos=b, frame = single, caption={Partially abstract specification, Ver. 2, \java{StudentRecord}}, basicstyle=\footnotesize]
class StudentRecord {
	{ ... } // Fields as in the Listing A.4
	
	//@ public invariant exam >= 0 && bonus >= 0 && passingGrade >= 0;
	//@ public invariant labs.length == 10;

	/*@
	@ public normal_behaviour
	@ requires_abs computeGradeR; 
	@ ensures_abs computeGradeE;
	@ assignable \nothing;
	@ def computeGradeR = bonus >= 0;
	@ def computeGradeE = (exam >= passingGrade ==> \result == exam + bonus) 
				&& (exam < passingGrade ==>\result == exam);
	@*/
    int computeGrade(){ ... }
	
	/*@
	@ public normal_behaviour
	@ requires_abs passedR;
	@ ensures_abs passedE1;
	@ ensures_abs passedE2;
	@ assignable \nothing;
	@ def passedR = true;
	@ def passedE1 = \result ==> exam + bonus >= passingGrade;
	@ def passedE2 = \result ==> (\forall int x; 0 <= x && x < 10; labs[x]); 
	@*/
	boolean passed() { ... }
}
\end{lstjavajml}
\bigskip

\begin{lstjavajml}[captionpos=b, frame = single, caption={Completely abstract specification, Ver. 2, \java{StudentRecord}}, basicstyle=\footnotesize]
class StudentRecord {
	{ ... } // Fields as in the Listing A.4
	
	//@ public invariant exam >= 0 && bonus >= 0 && passingGrade >= 0;
	//@ public invariant labs.length == 10;
	
	/*@
	@ public normal_behaviour
	@ requires_abs computeGradeR; 
	@ ensures_abs computeGradeE;
	@ assignable_abs computeGradeA;
	@ def computeGradeR = bonus >= 0;
	@ def computeGradeE = (exam >= passingGrade ==> \result == exam + bonus) 
				&& (exam < passingGrade ==>\result == exam);
	@ def computeGradeA = \nothing;
	@*/

    int computeGrade(){ ... }
	
	/*@
	@ public normal_behaviour
	@ requires_abs passedR;
	@ ensures_abs passedE1;
	@ ensures_abs passedE2;
	@ assignable_abs passedA;
	@ def passedR = true;
	@ def passedE1 = \result ==> exam + bonus >= passingGrade;
	@ def passedE2 = \result ==> (\forall int x; 0 <= x && x < 10; labs[x]); 
	@ def passedA = \nothing;
	@*/
	boolean passed() { ... }
}
\end{lstjavajml}
\bigskip
\vspace*{1cm}
\subsection{Version 3}\label{subsec:a-s-3}

\begin{lstjavajml}[captionpos=b, frame = single, caption={Concrete specification, Ver. 3, \java{StudentRecord}}, basicstyle=\footnotesize]
class StudentRecord {
	// exam result
	int exam;
	
	// minimum grade necessary to pass the exam
	int passingGrade;
	
	// completed labs
	boolean[] labs = new boolean[10];
	
	//@ public invariant exam >= 0 && passingGrade >= 0;
	//@ public invariant labs.length == 10;

	/*@
    @ public normal_behavior
    @ requires true;
    @ ensures \result == exam;
    @ assignable \nothing;
    @*/
    int computeGrade(){
        return exam;
    }
	
	/*@
	@ public normal_behavior
	@ ensures \result ==> exam >= passingGrade;
	@ ensures \result ==> (\forall int x; 0 <= x && x < 10; labs[x]);
	@ assignable \nothing;
	@*/
	boolean passed() {
		boolean enoughPoints = computeGrade() >= passingGrade;
		boolean allLabsDone = true;
		/*@ loop_invariant 0 <= i && i <= 10 
	    		 && (\forall int x; 0 <= x && x < i; 
			 			allLabsDone ==> labs[x]);
	   	@ assignable allLabsDone;
	   	@ decreases 10 - i;
	  	@*/
		for (int i = 0; i < 10; i++) {
	 	   allLabsDone = allLabsDone && labs[i];
		}
		return enoughPoints && allLabsDone;
	}
}
\end{lstjavajml}
\bigskip

\begin{lstjavajml}[captionpos=b, frame = single, caption={Partially abstract specification, Ver. 3, \java{StudentRecord}}, basicstyle=\footnotesize]
class StudentRecord {
	{ ... } // Fields as in the Listing A.7
	
	//@ public invariant exam >= 0 && passingGrade >= 0;
	//@ public invariant labs.length == 10;

	/*@
	@ public normal_behavior
	@ requires_abs computeGradeR; 
	@ ensures_abs computeGradeE;
	@ assignable \nothing;
	@ def computeGradeR = true;
	@ def computeGradeE = \result == exam;
	@*/
    int computeGrade(){ ... }
	
	/*@
	@ public normal_behavior
	@ requires_abs passedR;
	@ ensures_abs passedE1;
	@ ensures_abs passedE2;
	@ assignable \nothing;
	@ def passedR = true;
	@ def passedE1 = \result ==> exam >= passingGrade;
	@ def passedE2 = \result ==> (\forall int x; 0 <= x && x < 10; labs[x]); 
	@*/
	boolean passed() { ... }
}
\end{lstjavajml}

\begin{lstjavajml}[captionpos=b, frame = single, caption={Completely abstract specification, Ver. 3, \java{StudentRecord}}, basicstyle=\footnotesize]
class StudentRecord {
	{ ... } // Fields as in the Listing A.7
	
	//@ public invariant exam >= 0 && passingGrade >= 0;
	//@ public invariant labs.length == 10;

	/*@ public normal_behavior
	@ requires_abs computeGradeR; 
	@ ensures_abs computeGradeE;
	@ assignable_abs computeGradeA;
	@ def computeGradeR = true;
	@ def computeGradeE = \result == exam;
	@ def computeGradeA = \nothing;
	@*/
    int computeGrade(){ ... }
	
	/*@ public normal_behavior
	@ requires_abs passedR;
	@ ensures_abs passedE1;
	@ ensures_abs passedE2;
	@ assignable_abs passedA;
	@ def passedR = true;
	@ def passedE1 = \result ==> exam >= passingGrade;
	@ def passedE2 = \result ==> (\forall int x; 0 <= x && x < 10; labs[x]); 
	@ def passedA = \nothing;
	@*/
	boolean passed() { ... }
}
\end{lstjavajml}
\bigskip

\section{Code of the {\tt Account} example}\label{sec:a-account}
In this section of the appendix we provide the code of the \java{Account} example as it was used in our experiments from the Chapter Evaluation. For every version of the program we provide its implementation and concrete specification. The partially abstract specification, as well is the the completely abstract specification are omitted, because they can be straitforwardly derived from concrete specification following the same principles, that where illustrated in the previous section.
\subsection{Version 1}\label{subsec:a-a-1}
\begin{lstjavajml}[captionpos=b, frame = single, caption={Concrete specification, Ver.~1, \java{Account} and \java{Transaction}}, basicstyle=\footnotesize]
public class Account {
	/*@ accessible \inv:this.*; @*/
	
	int balance = 0;
	public boolean lock = false;
	
	/*
	 @ public normal_behavior
	 @ ensures (balance == \old(balance) + x) && \result; 
	 @ assignable balance;
	 @*/
	boolean update(int x) {
		balance = balance + x;
		return true;
	}

	/*@ 
	 @ public normal_behavior
	 @ ensures (balance == \old(balance) - x) && \result;
	 @ assignable balance;
	 @*/
	boolean undoUpdate(int x) {
		balance = balance - x;
		return true;
	}
	
	/*@
	 @ public normal_behavior
	 @ ensures \result == this.lock;
	 @*/
	boolean /*@ pure @*/ isLocked() {
		return lock;
	}
}

public class Transaction {
	/*@ accessible \inv:this.*; @*/

	/*@ public normal_behavior
	  requires destination != null && source != null && \invariant_for(source) 
	  		&& \invariant_for(destination);
	  requires source != destination;
	  ensures \result ==> (\old(destination.balance) + amount == destination.balance);
	  ensures \result ==> (\old(source.balance) - amount == source.balance);
	  assignable \everything;
	 @*/
	public boolean transfer(Account source, Account destination, int amount) {
		if (source.balance < 0) amount = -1;
		if (destination.isLocked()) amount = -1;
		if (source.isLocked()) amount = -1;
		
		int take;
		int give;
		if (amount != -1) { take = amount * -1; give = amount;} 
	
		if (amount <= 0) {
			return false;
		}
		if (!source.update(take)) {
			return false;
		}
		if (!destination.update(give)) {
			source.undoUpdate(take);
			return false;
		}
		return true;	
	}
}
\end{lstjavajml}
\bigskip

\subsection{Version 2}\label{subsec:a-a-2}
\begin{lstjavajml}[captionpos=b, frame = single, caption={Concrete specification, Ver.~2, \java{Account} and \java{Transaction}}, basicstyle=\footnotesize]
public class Account {
	/*@ accessible \inv:this.*; @*/
	
	int balance = 0;
	public boolean lock = false;
	
	/*@
	 @ public normal_behavior
	 @ ensures \result; 
	 @ assignable balance;
	 @*/
	boolean update(int x) {
		balance = balance + x;
		return true;
	}

	/*@ 
	 @ public normal_behavior
	 @ ensures \result;
	 @ assignable balance;
	 @*/
	boolean undoUpdate(int x) {
		balance = balance - x;
		return true;
	}
	
	/*@
	 @ public normal_behavior
	 @ ensures \result == this.lock;
	 @*/
	boolean /*@ pure @*/ isLocked() {
		return lock;
	}	
}

public class Transaction {
	/*@ accessible \inv:this.*; @*/

	/*@ public normal_behavior
	  requires destination != null && source != null && \invariant_for(source) 
	  		&& \invariant_for(destination);
	  requires source != destination;
	  ensures true;
	  ensures (amount <= 0) ==> !\result;
	  assignable \everything;
	 @*/
	public boolean transfer(Account source, Account destination, int amount) {
		if (source.balance < 0) amount = -1;
		if (destination.isLocked()) amount = -1;
		if (source.isLocked()) amount = -1;
		
		int take;
		int give;
		if (amount != -1) { take = amount * -1; give = amount;} 
	
		if (amount <= 0) {
			return false;
		}
		if (!source.update(take)) {
			return false;
		}
		if (!destination.update(give)) {
			source.undoUpdate(take);
			return false;
		}
		return true;	
	}
}
\end{lstjavajml}
\bigskip
\vspace*{1cm}
\subsection{Version 3}\label{subsec:a-a-3}
\begin{lstjavajml}[captionpos=b, frame = single, caption={Concrete specification, Ver.~3, \java{Account} and \java{Transaction}}, basicstyle=\footnotesize]
public class Account {
	/*@ accessible \inv:this.*; @*/
	
	final int OVERDRAFT_LIMIT = 0;
	
	//@ public invariant balance >= OVERDRAFT_LIMIT;
	int balance = 0;
	
	public boolean lock = false;

	/*@ 
	 @ public normal_behavior
	 @ ensures (!\result ==> balance == \old(balance)) 
	 @   && (\result ==> balance == \old(balance) + x); 
	 @ assignable balance;
	 @*/
	boolean update(int x) {
		int newBalance = balance + x;
		if (newBalance < OVERDRAFT_LIMIT)
			return false;
		balance = newBalance;
		return true;
	}

	/*@ 
	 @ public normal_behavior
	 @ ensures (!\result ==> balance == \old(balance)) 
	 @   && (\result ==> balance == \old(balance) - x);
	 @ assignable balance;
	 @*/
	boolean undoUpdate(int x) {
		int newBalance = balance - x;
		if (newBalance < OVERDRAFT_LIMIT)
			return false;
		balance = newBalance;
		return true;
	}
	
	/*@
	 @ public normal_behavior
	 @ ensures \result == this.lock;
	 @*/
	boolean /*@ pure @*/ isLocked() {
		return lock;
	}	
}

public class Transaction {
	/*@ accessible \inv:this.*; @*/

	/*@ public normal_behavior
	  requires destination != null && source != null && \invariant_for(source) 
	  		&& \invariant_for(destination);
	  requires source != destination;
	  ensures \result ==> (\old(destination.balance) + amount == destination.balance);
	  ensures \result ==> (\old(source.balance) - amount == source.balance);
	  assignable \everything;
	 @*/
	public boolean transfer(Account source, Account destination, int amount) {
		if (source.balance < 0) amount = -1;
		if (destination.isLocked()) amount = -1;
		if (source.isLocked()) amount = -1;
		
		int take;
		int give;
		if (amount != -1) { take = amount * -1; give = amount;} 
	
		if (amount <= 0) {
			return false;
		}
		if (!source.update(take)) {
			return false;
		}
		if (!destination.update(give)) {
			source.undoUpdate(take);
			return false;
		}
		return true;	
	}
}
\end{lstjavajml}
\bigskip

\subsection{Version 4}\label{subsec:a-a-4}
\begin{lstjavajml}[captionpos=b, frame = single, caption={Concrete specification, Ver.~4, \java{Account} and \java{Transaction}}, basicstyle=\footnotesize]
public class Account {
	/*@ accessible \inv:this.*; @*/
	
	final int OVERDRAFT_LIMIT = 0;
	final static int DAILY_LIMIT = -1000;
	
	//@ public invariant balance >= OVERDRAFT_LIMIT;
	int balance = 0;
	
	//@ invariant withdraw >= DAILY_LIMIT;
	int withdraw = 0;
	
	public boolean lock = false;
	
	/*@ public normal_behavior
	 @ ensures (!\result ==> balance == \old(balance)) 
	 @   && (\result ==> balance == \old(balance) + x)
	 @	 && (!\result ==> withdraw == \old(withdraw)) 
	 @   && (\result ==> withdraw <= \old(withdraw));
	 @ assignable balance;
	 @*/
	boolean update(int x) {
		int newWithdraw = withdraw;
		if (x < 0)  {
			newWithdraw += x;
			if (newWithdraw < DAILY_LIMIT) 
				return false;
		}
		int newBalance = balance + x;
		if (newBalance < OVERDRAFT_LIMIT)
			return false;
		balance = newBalance;
		withdraw = newWithdraw;
		return true;
	}

	/*@ public normal_behavior
	 @ ensures (!\result ==> balance == \old(balance)) 
	 @   && (\result ==> balance == \old(balance) - x)
	 @   && (!\result ==> withdraw == \old(withdraw)) 
	 @   && (\result ==> withdraw >= \old(withdraw));
	 @ assignable balance;
	 @*/
	boolean undoUpdate(int x) {
		int newWithdraw = withdraw;
		if (x < 0)  {
			newWithdraw -= x;
			if (newWithdraw < DAILY_LIMIT) 
				return false;
		}
		int newBalance = balance - x;
		if (newBalance < OVERDRAFT_LIMIT)
			return false;
		balance = newBalance;
		withdraw = newWithdraw;	
	
		return true;
	}
	
	/*@
	 @ public normal_behavior
	 @ ensures \result == this.lock;
	 @*/
	boolean /*@ pure @*/ isLocked() {
		return lock;
	}
}	
public class Transaction {
	/*@ accessible \inv:this.*; @*/

	/*@ public normal_behavior
	  requires destination != null && source != null && \invariant_for(source) 
	  		&& \invariant_for(destination);
	  requires source != destination;
	  ensures \result ==> (\old(destination.balance) + amount == destination.balance);
	  ensures \result ==> (\old(source.balance) - amount == source.balance);
	  assignable \everything;
	 @*/
	public boolean transfer(Account source, Account destination, int amount) {
		if (source.balance < 0) amount = -1;
		if (destination.isLocked()) amount = -1;
		if (source.isLocked()) amount = -1;
		
		int take;
		int give;
		if (amount != -1) { take = amount * -1; give = amount;} 
	
		if (amount <= 0) {
			return false;
		}
		if (!source.update(take)) {
			return false;
		}
		if (!destination.update(give)) {
			source.undoUpdate(take);
			return false;
		}
		return true;	
	}
}

\end{lstjavajml}
\bigskip

\subsection{Version 5}\label{subsec:a-a-5}
\begin{lstjavajml}[captionpos=b, frame = single, caption={Concrete specification, Ver.~5, \java{Account} and \java{Transaction}}, basicstyle=\footnotesize]

public class Account {
	/*@ accessible \inv:this.*; @*/
	
	final int OVERDRAFT_LIMIT = 0;
	final int FEE = 1;
	
	//@ public invariant balance >= OVERDRAFT_LIMIT;
	//@ public invariant FEE >= 0;
	
	int balance = 0;
	public boolean lock = false;
	
	/*@ public normal_behavior
	 @ ensures (!\result ==> balance == \old(balance)) 
	 @   && (\result ==> balance == \old(balance) + x - FEE); 
	 @ assignable balance;
	 @*/
	boolean update(int x) {
		int newBalance = balance + x - FEE;
		if (newBalance < OVERDRAFT_LIMIT)
			return false;
		balance = newBalance;
		return true;
	}

	/*@ public normal_behavior
	 @  ensures (!\result ==> balance == \old(balance)) 
	 @   && (\result ==> balance == \old(balance) - x + FEE);
	 @  assignable balance;
	 @*/
	boolean undoUpdate(int x) {
		int newBalance = balance - x + FEE;
		if (newBalance < OVERDRAFT_LIMIT)
			return false;
		balance = newBalance;
		return true;
	}
	
	/*@
	 @ public normal_behavior
	 @ ensures \result == this.lock;
	 @*/
	boolean /*@ pure @*/ isLocked() {
		return lock;
	}
}

public class Transaction {
	/*@ accessible \inv:this.*; @*/

	/*@ public normal_behavior
	  requires destination != null && source != null && \invariant_for(source) 
	  		&& \invariant_for(destination);
	  requires source != destination;
	  ensures \result ==> (\old(destination.balance) + amount >= destination.balance);
	  ensures \result ==> (\old(source.balance) - amount >= source.balance);
	  assignable \everything;
	 @*/
	public boolean transfer(Account source, Account destination, int amount) {
		if (source.balance < 0) amount = -1;
		if (destination.isLocked()) amount = -1;
		if (source.isLocked()) amount = -1;
		
		int take;
		int give;
		if (amount != -1) { take = amount * -1; give = amount;} 
	
		if (amount <= 0) {
			return false;
		}
		if (!source.update(take)) {
			return false;
		}
		if (!destination.update(give)) {
			source.undoUpdate(take);
			return false;
		}
		return true;	
	}
}
\end{lstjavajml}

\vspace*{2cm}

\section{Code of the {\tt Log} class}\label{sec:a-log}
In this section of the appendix we provide the code for the three versions of the \java{Log} class as it was used in our additional experiments in Chapter \ref{ch:evaluation}. For every version of the program we provide its implementation and concrete specification. The completely abstract specification can be straitforwardly derived from concrete specification and the partially abstract specification is not relevant, because the assignable clauses in this example differ among the versions.

\subsection{Ring}
\begin{lstjavajml}[captionpos=b, frame = single, caption={Concrete specification, Ring \java{Log}}, basicstyle=\footnotesize]
public class Log {
	/*@ accessible \inv:this.*; @*/

    //@ public invariant logRecord.length > 0;
    protected /*@ spec_public @*/ int[] logRecord;

    //@ public invariant last >= -1 && last < logRecord.length;
    protected /*@ spec_public @*/ int last;
   
    public Log(int size) {
		this.logRecord = new int[size];
		last = -1;
    }
   
    /*@ public normal_behavior
      @ requires true;
      @ ensures (\result == (last == logRecord.length - 1 ? 0 : last + 1));
      @ assignable \nothing;
      @*/
    protected /*@ pure @*/ int rotateLog() {
		return (last + 1) \% logRecord.length;
    }

    /*@ public normal_behavior
      @ requires   true;
      @ ensures    (last == (\old(last) == logRecord.length - 1 ? 0 : \old(last) + 1)) && 
      @             (logRecord[last] == bal);
      @ assignable logRecord[(last + 1) \% logRecord.length], last;
      @*/
    public void add(int bal) {
		last = rotateLog();
        	logRecord[last] = bal;
    }
}
\end{lstjavajml}
\bigskip

\subsection{Empty}
\begin{lstjavajml}[captionpos=b, frame = single, caption={Concrete specification, Empty \java{Log}}, basicstyle=\footnotesize]
public class Log {
	/*@ accessible \inv:this.*; @*/
		
	//@ public invariant logRecord.length > 0;
     protected /*@ spec_public @*/ int[] logRecord;

	//@ public invariant last >=-1 && last < logRecord.length;
	protected /*@ spec_public @*/ int last;
	
	public Log(int size) {
		this.logRecord = new int[size];
		last = -1;
	}
   
    /*@ public normal_behavior
      @ requires true;
      @ ensures  (\result == ((logRecord.length-1 == last) ? 0 : last + 1)) &&
      @     ((logRecord.length-1 == last) ==>  
      @			(\forall int i; i>=0 && i<logRecord.length; logRecord[i] == 0));
      @ assignable logRecord[*];
      @*/
	public int rotateLog() {
		if (last == logRecord.length - 1) {
			// empty array
	    	  /*@ loop_invariant
	      	@   i>=0 && i<=logRecord.length &&
	      	@   	(\forall int j; j>=0 && j<i; logRecord[j] == 0);
	      	@ decreases logRecord.length - i;
	      	@ assignable logRecord[*];
	      	@*/
			for (int i = 0; i<logRecord.length; i++) {
				logRecord[i] = 0;		
	   		}
	    		return 0;
		} else {
	   		return last + 1;
		}
	}

    /*@ public normal_behavior
      @ requires true;
      @ ensures last == (\old(last) == logRecord.length  - 1 ? 0 : \old(last) + 1) && 
      @         logRecord[last] == bal;
      @ assignable logRecord[*], last;
      @*/
    public void add(int bal) {
		last = rotateLog();
        	logRecord[last] = bal;
    }
}
\end{lstjavajml}
\bigskip

\subsection{Replace}
\begin{lstjavajml}[captionpos=b, frame = single, caption={Concrete specification, Replace \java{Log}}, basicstyle=\footnotesize]
public class Log {
	/*@ accessible \inv:this.*; @*/
	
    //@ public invariant logRecord.length > 0;
    protected /*@ spec_public @*/ int[] logRecord;

    //@ public invariant last >=-1 && last < logRecord.length;
    protected /*@ spec_public @*/ int last;
   
    public Log(int size) {
		this.logRecord = new int[size];
		last = -1;
    }
   
    /*@ public normal_behavior
      @ requires true;
      @ ensures  (\result == ((logRecord.length - 1 == \old(last)) ? 0 : \old(last) + 1)) &&
      @      ((logRecord.length - 1 == \old(last)) ==> 
      @             ( \fresh(logRecord) &&
      @               (\forall int i; i>=0 && i<logRecord.length; logRecord[i] == 0) &&
      @               logRecord.length == \old(logRecord.length))) &&
      @      ((logRecord.length - 1 > \old(last)) ==> logRecord == \old(logRecord));
      @ assignable logRecord;
      @*/
    public int rotateLog() {
		if (last == logRecord.length - 1) {
	    		logRecord = new int[logRecord.length];
	    		return 0;
		} else {
	    		return last + 1;
		}
    }

    /*@ public normal_behavior
      @ requires   true;
      @ ensures    last == (\old(last) == \old(logRecord.length) - 1 ? 0 : \old(last) + 1) && 
      @                logRecord[last] == bal;
      @ assignable logRecord, last, logRecord[last + 1];
      @*/
    public void add(int bal) {
		last = rotateLog();
        	logRecord[last] = bal;
    }
}
\end{lstjavajml}
\bigskip

\end{document}